\documentclass{article}
\usepackage{epsfig}
\usepackage{a4,amsmath,amsthm,amsfonts,amscd,eucal,latexsym,amssymb}

\def\cA{{\cal A}}
\def\cB{{\cal B}}

\def\cH{{\cal H}}

\def\cR{{\cal R}}

\def\cT{{\cal T}}

\def\cX{{\cal X}}

\def\bC{{\mathbb C}}

\def\bN{{\mathbb N}}
\def\bR{{\mathbb R}}

\def\a{\alpha}
\def\b{\beta}

\def\m{\mu}

\def\p{\pi}
\def\r{\rho}

\def\o{\omega}
\def\O{\Omega}
\def\fA{{\mathfrak A}}

 \newtheorem{Definition}{Definition}[section]
 \newtheorem{Theorem}[Definition]{Theorem}
 \newtheorem{Proposition}[Definition]{Proposition}
 \newtheorem{Lemma}[Definition]{Lemma}
 \newtheorem{Corollary}[Definition]{Corollary}

\def\qed{${}$ \quad ${}$ \hfill $\Box$ \\[6pt]}

\newsymbol\bt 1202  

\newcommand{\Bof}[1]{{B(#1)}}
\newcommand{\Bofcc}{\Bof{\bC^2}}
\newcommand{\eins}{{\rm 1\!\!l}}
\let\idty\eins

\begin{document}
\noindent
\begin{center}
{ \Large \bf Distillability and Positivity of Partial Transposes \\[6pt]
in General Quantum Field Systems}
\\[30pt]
{\large \sc
Rainer Verch${}^{(1)}$ and Reinhard F.\ Werner${}^{(2)}$}
\end{center}
${}$\\
                 ${}^{(1)}$\,Max-Planck-Institut for
                 Mathematics in the Sciences,\\
                 Inselstr.\ 22,
                 D-04103 Leipzig, Germany ---
                 e-mail: verch$@$mis.mpg.de
${}$\\[4pt]  ${}^{(2)}$\,Institut f.\ Mathematische Physik, TU Braunschweig,\\
 Mendelssohnstr.\ 3, D-38106 Braunschweig, Germany --- e-mail: R.Werner@tu-bs.de
\\[22pt]
${}$ \hfill {\it Dedicated to Detlev Buchholz on the occasion of his sixtieth birthday}
${}$\\[26pt]
{\small {\bf Abstract. }  Criteria for distillability, and the
property of having a positive partial transpose, are introduced
for states of general bipartite quantum systems. The framework is
sufficiently general to include systems with an infinite number of
degrees of freedom, including quantum fields.  We show  that a
large number of states in  relativistic quantum field
theory, including the vacuum state and thermal equilibrium states,
are distillable over subsystems separated by arbitrary spacelike
distances. These results apply to any quantum field model. It will
also be shown that these results can be generalized to quantum
fields in curved spacetime, leading to the conclusion that there
is a large number of quantum field states which are distillable
over subsystems separated by an event horizon. } \noindent 

\section{Introduction}
In the present work we investigate entanglement criteria for
quantum systems with infinitely many degrees of freedom, paying
particular attention to relativistic quantum field theory.

The specification and characterization of entanglement in
quantum systems is a primary issue in quantum information theory
(see \cite{Keyl} for a recent review of quantum information
theory). Entanglement frequently appears as a resource for typical
quantum information tasks, in particular for teleportation
\cite{?c}, key distribution \cite{Ekert}, and quantum
computation \cite{one-way}. Ideally these processes use bipartite
entanglement in the form of maximally entangled states, such as
the singlet state of two spin-$1/2$ particles. But less entangled
sources can sometimes be converted to such maximally entangled
ones by a some ``distillation process'' using only local quantum
operations and classical communication \cite{Popescu94,BDSW}.
States for which this is possible are called ``distillable'', and
this property is the strongest entanglement property for generic
states (as opposed to special parameterized families). Indeed, it
is stronger than merely being entangled, where a state is called
entangled if it cannot be written as a mixture of uncorrelated
product states. The existence of non-distillable entangled states
(also called ``bound entangled states'') was first shown in
\cite{H3}.

For a given state it is often not easy  to decide to which class
it belongs. A very efficient criterion is obtained from studying
the partial transpose of the density operator, and asking whether
it is a positive operator. In this case the state is called a ppt
state, and an npt state otherwise. Originally, the npt property
was established by Peres \cite{Per} as a sufficient condition
for entanglement, and was subsequently shown to be also sufficient
for low dimensional systems \cite{Hiii,Wor} and some highly
symmetric systems \cite{Vollb}.

It turns out that ppt states cannot be distilled, so the existence
of bound entangled states shows that the ppt condition is a much
tighter fit for non-distillability than for mere separability. In
fact, it is one of the major open problems \cite{problem2} to
decide whether there is equivalence, i.e., whether all npt states
are distillable. There have been indications that this conjecture
might fail for bipartite quantum systems having finitely and
sufficiently many discrete degrees of freedom \cite{DiVSSTT,DCLB}.
On the other hand, for bipartite quantum systems having finitely
many continuous degrees of freedom (such as harmonic oscillators)
it was found that Gaussian states which are npt  are  also
distillable (about this and related results, cf.\ \cite{GKDZCL}
and refs.\ cited there).

While this brief recapitulation of results documents that the
distinction between entangled, npt and distillable states is a
subtle business already in the case of quantum systems with
finitely many degrees of freedom, we should now like to point out
that the study of entanglement is also a longstanding issue in
general quantum field theory.
Already before the advent of quantum
information theory, the extent to which Bell-inequalities are
violated has been investigated in several articles by Landau
\cite{Lan1,Lan2} and by Summers and Werner
\cite{SumWer87a,SumWer,SJS-RMP2}. In fact, the studies
\cite{SumWer87a,SumWer} motivated the modern concept of separable
states (then called ``classically correlated'' \cite{Wern89}) and
raised the question of the connection between separability and
Bell's inequalities. More recently, there has been a renewed
interest \cite{RRS,HalCl,HNarn,BeckGottNielPres,EggeSchlWer,PerTer} in the
connection between ``locality'' as used in quantum information
theory on the one hand, and in quantum field theory on the other.
However, for some of the relevant questions, like distillability, the usual framework
of quantum information theory, mainly focussing on systems with
finite dimensional Hilbert spaces, is just not rich enough. This
lack, which is also serious for the connections between
entanglement theory and statistical mechanics of infinite systems,
is addressed in the first part of our paper.

In particular, we extend the notions of separability and
distillability for the general bipartite situations found in
systems which have infinitely many degrees of freedom, and which cannot
be expressed in terms of the tensor product of Hilbert spaces.
These generalizations are fairly straightforward. Less obvious is
our generalization of the notion of states with positive partial
transpose, since the operation ``partial transposition'' itself
becomes meaningless. Of course, we also establish the usual
implications between these generalized concepts.

It turns out that 1-distillability of a state follows from the
Reeh-Schlieder property, which has been thoroughly investigated
for quantum field theoretical systems. After establishing this
connection, we can therefore bring to bear known results from
quantum field theory to draw some new conclusions about the
non-classical nature of vacuum fluctuations. In particular, the
vacuum is 1-distillable, even when Alice and Bob operate in
arbitrarily small spacetime regions, and arbitrarily far apart in
a Minkowski spacetime.  Such a form of distillability can then
also be deduced to hold for a very large (in a suitable sense,
dense) class of quantum field states, including thermal equilibrium states.
We comment on related results in \cite{HalCl,Jae3} and \cite{RRS} in the
remarks following Thm.\ \ref{QFTdistill}.
Furthermore, we generalize the distillability 
result to free quantum fields on curved spacetimes. We also point
out that this entails distillability of a large class of quantum
field states over subsystems which may be separated by an event
horizon in spacetime, inhibiting two-way classical communication
between the system parts, and we will discuss what this means for
the distillability concept.

\section{General Bipartite Quantum Systems}
\setcounter{equation}{0}
 The bipartite quantum systems arising in quantum field theory are
systems of infinitely many degrees of freedom. In contrast, the
typical descriptions of concepts and results of quantum
information theory are for quantum systems described in finite
dimensional Hilbert spaces. In this section we describe the basic
mathematical structures needed to describe systems of infinitely
many degrees of freedom and, in particular, bipartite systems in
that context.

For the transition to infinitely many degrees of freedom it does
not suffice to consider Hilbert spaces of infinite dimension: this
level of complexity is already needed for a single harmonic
oscillator. The key idea allowing the transition to infinitely
many oscillators is to look at the observable algebra of the
system, which is then no longer the algebra of all bounded
operators on a Hilbert space, but a more general operator algebra.
This operator  algebraic approach to large quantum systems has
proved useful in both quantum field theory and quantum statistical
mechanics \cite{BratRob1,BratRob2,Emch,Haag,Sew}.

For many questions we discuss, it suffices to take the observable
algebra $\cR$ as a general C*-algebra: this
is defined as an algebra with an adjoint operation $X \mapsto X^*$
on the algebra elements $X$ and also with a norm with respect to
which it is complete and which satisfies $||X^*X||=||X||^2$.
In practically all applications, $\cR$ is given in a Hilbert space representation,
so that it is usually no restriction of generality to think
of $\cR$ as a
norm-closed and adjoint-closed subalgebra of the algebra $\Bof\cH$ of all
bounded linear operators on a Hilbert space $\cH$.
We should emphasize, though, that $\cR$ is usally really a proper
subalgebra of $\Bof\cH$, and also its weak closure (in the sense of 
convergence of expectation values) $\overline{\cR}$ will typically
be a proper subalgebra of $\Bof\cH$. This is of particular importance
in the present context when we consider $C^*$-algebras of local observables
in relativistic quantum field theories: These are proper subalgebras
of some $\Bof\cH$ which don't contain any finite-dimensional projection
(in technical terms, the von Neumann algebras arising as their weak
closures are purely infinite, cf.\ Sec.\ 2.7 in \cite{BratRob1}).
 Therefore, the properties of
these algebras are fundamentally different from those of the full $\Bof\cH$;
in particular, arguments previously developed in quantum information theory
for finite dimensional systems modelled on $B(\bC^m) \otimes B(\bC^n)$ 
 are typically based on the use
of finite-dimensional projections and thus they can usually not simply be
generalized to the quantum field theoretical case. 

We only consider algebras with unit element $\eins$. For some
questions we will consider a special type of such algebras, called von
Neumann algebras, about which we collect some basic facts later.
In any case, the ``observables'' are specified as selfadjoint
elements of the algebra, or, more generally as measures (POVMs)
with values in the positive elements of $\cR$.

Discussions of entanglement always refer to distinguished
subsystems of a given quantum system. Subsystems are specified as
subalgebras of the total observable algebra. For a bipartite
system we must specify two subsystems with  the crucial property
that every observable of one subsystem can be measured jointly
with every observable of the other, which is equivalent to saying
that the observable algebras commute elementwise. Hence we arrive
at

\begin{Definition} A (generalized) {\bf bipartite system}, usually denoted by
$(\cA,\cB)\subset\cR$, is a pair of C*-subalgebras $\cA,\cB$ of a
larger C*-algebra $\cR$, called the ambient algebra of the system,
such that the identity is contained in both algebras, and all
elements of $A\in\cA$ and $B\in\cB$ commute.
\end{Definition}

Thinking of typical situations in quantum information theory,
$\cA$ corresponds to the observables controlled by `Alice' and
$\cB$ to the observables controlled by `Bob'. The ambient algebra
$\cR$ will not play an important role for the concepts we define.
For most purposes it is equivalent to choose $\cR$ either
``minimal'', i.e., as the smallest C*-subalgebra containing both
$\cA$ and $\cB$, or else ``maximal'' as $\Bof\cH$, the algebra of
all bounded operators on the Hilbert space $\cH$ on which all the
operators under consideration are taken to operate.

The standard quantum mechanical example of a bipartite situation
is given by  the tensor product $\cH = \cH_{\rm A} \otimes
\cH_{\rm B}$ of two Hilbert-spaces $\cH_{\rm A}$ and $\cH_{\rm
B}$, with the observable algebras of Alice and Bob defined as
 $\cR = \Bof\cH$, $\cA = \Bof{\cH_{\rm A}} \otimes \eins$,
$\cB = \eins \otimes \Bof{\cH_{\rm B}}$.

Note that in this example both algebras $\cA,\cB$ are of the
form $B(\widetilde{\cH})$ (for suitable Hilbert-space $\widetilde{\cH}$),
and as mentioned above, this will not be the case any more when $\cA$ 
and $\cB$ correspond to algebras of local observables in quantum field
theory. Furthermore,
  if we do not want to impose unnecessary
algebraic restrictions on the subsystems, we must envisage more
general compositions than of tensor product form, 
too.  Such systems arise naturally in quantum
field theory, for tangent spacetime regions \cite{SumWer}, but also
if we want to describe a state of an infinite collection of
singlet pairs, and other ``infinitely entangled'' situations
\cite{KSW}.

A {\it state} on a C*-algebra $\cR$ is a linear functional $\o :
\cR \to \bC$, which can be interpreted as an expectation value
functional, i.e.,  which is positive ($\o(A) \ge 0$ for $A \ge
0$), and normalized ($\o(\eins) = 1$). When
$\cR\subset\cB(\cH)$, i.e., when we consider a particular {\it
representation} of all algebras involved as algebras of operators,
we can consider the special class of states of the
form\begin{equation}\label{e:normal}
    \o(A) = {\rm Tr}(\r_{\o}A) \quad {\rm for\ all}\ A \in \cR\,,
\end{equation}
for some positive trace class operator $\r_{\o}$, called the
density operator of $\r$. Such states are called {\it normal}
(with respect to the representation). As usual, for $A = A^*$
representing an observable, the value $\o(A)$ is the expectation
value of the observable $A$ in the state $\o$. A {\it bipartite
state} is simply a state on the ambient algebra of a bipartite
system. Since every state on the minimal ambient algebra can be
extended to a state on the maximal algebra, this notion does not
intrinsically depend on the choice of ambient algebra.

A bipartite state $\o$ is a {\it product state} if
$\o(AB)=\o(A)\o(B)$ for all $A\in\cA$ and $B\in\cB$. Similarly,
$\o$ is called {\it separable}, if it is the weak
limit\footnote{This means that $\lim_\alpha\o_\alpha(X)=\o(X)$,
for all $X\in\cR$} of states $\o_\a$, each of which is a convex
combination of product states.

\section{Positivity of Partial Transpose (ppt)}

Consider again the standard situation in quantum information
theory, where all Hilbert spaces are finite dimensional, and a
bipartite system with Hilbert space
 $\cH = \cH_{\rm A}\otimes\cH_{\rm B}$.
Then we can define the {\it partial transpose} of a state $\o$, or
equivalently, its density matrix with $\r_\o$ with $\o(\,.\,) =
{\rm Tr}(\r_{\o}\,.\,)$, by introducing orthonormal bases
$\{|e^{({\rm A})}_k\rangle\}$ in $\cH_{\rm A}$ and $\{|e^{({\rm
B})}_{\ell}\rangle\}$ in $\cH_{\rm B}$ for each of the Hilbert
spaces, and swapping the matrix indices belonging to one of the
factors, say the first, so that
\begin{equation}\label{ptrans}
  \langle e^{({\rm A})}_k \otimes e^{({\rm B})}_{\ell}|
        \r_{\o}^{T_1}| e^{({\rm A})}_m \otimes e^{({\rm B})}_n \rangle
 =\langle e^{({\rm A})}_m \otimes e^{({\rm B})}_{\ell}|
     \r_{\o}| e^{({\rm A})}_k \otimes e^{({\rm B})}_n\rangle\;.
\end{equation}
Then it is easy to see that in general $\r_o\geq0$ does not imply
$\r_{\o}^{T_1}\geq0$, i.e., the partial transpose operation is not
completely positive. On the other hand, if $\o$ is separable, then
$\r_{\o}^{T_1}\geq0$. More generally, we say that $\o$ is a {\it
ppt-state} when this is the case. As we just noted, the ppt
property is necessary for separability, and also sufficient in low
dimensions ($2\otimes2$ and $2\otimes3$), which is known as the
Peres-Horodecki criterion for separability \cite{Per}.

It is important to note that while the definition of the partial
transpose depends on the choice of bases, the ppt-condition does
not: different partial transposes are linked by a unitary
transformation and so have the same spectrum. In the more involved
context of general bipartite systems, we will follow a similar
approach by defining a ppt property without even introducing an
object which one might call the `partial transpose' of the given
state, and which would in any case be highly dependent on further
special choices.

\begin{Definition}\label{def:ppt}
We say that a state $\o$ on a bipartite system
$(\cA,\cB)\subset\cR$ has the {\bf ppt} property if
 for any choice of finitely many
$A_1,\ldots,A_k \in \cA$, and $B_1,\ldots,B_k \in \cB$, one has
$$ \sum_{\a,\b} \o(A_{\b}A^*_{\a}B_{\a}^* B_{\b}) \ge 0\,.$$
\end{Definition}

Clearly, this definition is independent of the choice of ambient
algebra $\cR$, since only expectations of the form $\o(AB)$ enter.
It is also symmetrical with respect to the exchange of $\cA$ and
$\cB$ (just exchange $A_\alpha$ and $B_\beta^*$, with concomitant
changes).

Our first task is to show that this notion of ppt coincides with
that given by Peres \cite{Per} in the case of finite-dimensional Hilbert
spaces. We show this by looking more generally at situations in
which there is a candidate for the role of the ``partial transpose
of $\o$''.

\begin{Proposition}
Let $(\cA,\cB)\subset\Bof\cH$ be a bipartite system, and let
$\theta$ be an anti-unitary operator on $\cH$ such that the
algebra $\widetilde\cB\equiv\theta^*\cB\theta$ commutes
elementwise with $\cA$.\\
 (1) Suppose that $\widetilde\o$ is a state
on $\Bof\cH$ such that
\begin{equation}\label{omegapT}
    \widetilde\o(A\widetilde B)=\o(A\theta \widetilde B^*\theta^*)
\end{equation}
for $A\in\cA$ and $\widetilde B\in\widetilde\cB$.  Then
$\omega$ is ppt. \\
 (2) In particular, if $\cA$, $\cB$ are finite
dimensional matrix algebras, Definition~\ref{def:ppt} is
equivalent to the positivity of the partial transpose in the sense
of Eq.~(\ref{ptrans}).
\end{Proposition}

Note that the star on the right hand side of Eq.~(\ref{omegapT})
is necessary so the whole equation becomes linear in $\widetilde
B$. When $\theta$ is complex conjugation in some basis, $X\mapsto
\theta^* X^*\theta$ is exactly the matrix transpose in that basis.
This proves the second part of the Proposition: if $\cA,\cB$ are
matrix algebras, we can identify $\cB$ with the algebra of all
transposed matrices $\theta^*\cB\theta$, and with this
identification Eq.~(\ref{omegapT}) defines a linear functional on
$\cA\otimes\cB$, which is just the partial transpose of $\omega$.
The only issue for the ppt property in both formulations is indeed
whether this functional is positive, i.e., a state.

{\it Proof. } Let $A_1,\ldots,A_k \in \cA$, and $B_1,\ldots,B_k
\in \cB$ be as in Definition~\ref{def:ppt}, and introduce
$\widetilde B_\a=\theta^*B^*_\a\theta$, so that also
 $B_\a=\theta\widetilde B^*_\a\theta^*$. Then
\begin{eqnarray}\label{hjk}
 \sum_{\a,\b}\o\left( A_{\b}A^*_{\a}\ B_{\a}^* B_{\b}\right)
   &=&\sum_{\a,\b}\o\left( A_{\b}A^*_{\a}\
          \theta \widetilde B_{\a} \theta^*
           \theta\widetilde B_{\b}^* \theta^* \right)\nonumber\\
   &=&\sum_{\a,\b}\o\left( A_{\b}A^*_{\a}\
          \theta \widetilde B_{\a}
          \widetilde B_{\b}^* \theta^* \right)\nonumber\\
   &=&\sum_{\a,\b}\o\left( A_{\b}A^*_{\a}\
          \theta (\widetilde B_{\b}\widetilde B_{\a}^*)^*
            \theta^* \right)\nonumber\\
   &=&\sum_{\a,\b}\widetilde\o\left( A_{\b}A^*_{\a}\
          \widetilde B_{\b}\widetilde B_{\a}^*
            \right)\nonumber\\
   &=&\widetilde\o(XX^*)\;,
\end{eqnarray}
with $X=\sum_\alpha A_\alpha \widetilde B_\alpha$. Clearly, when
$\widetilde\o$ is a state, this is positive. \qed

Another consistency check is the following.
\begin{Lemma}\label{sep2ppt}
Also for general bipartite systems, separable states are ppt.
\end{Lemma}
{\it Proof. } Obviously,the ppt property is preserved under weak
limits and convex combinations. By definition, each separable
state arises by such operations from product states. Hence it is
enough to show that each product state on $\cR$ is ppt. If
$\o(AB)=\o(A)\o(B)$ is a product state, and $A_1,\ldots,A_k \in
\cA$, and $B_1,\ldots,B_k \in \cB$ we introduce the $k\times
k$-matrices $M_{\b\a}=\o(A_\b A_\a^*)$ and
$N_{\a\b}=\o(B_\a^*B_\b)$. What we have to show according to
Definition~\ref{def:ppt} is that ${\rm tr}(MN)\geq0$. But this is
clear from the observation that $M$ and $N$ are obviously positive
semi-definite. \qed

Therefore the set of states which are {\it not} ppt across $\cA$
and $\cB$ (the ``npt-states'') forms a subset of the class of
entangled states. As is well-known already for low dimensional
examples (larger than $3\otimes3$-dimensional systems) the
converse of this Lemma fails.

We add another result, an apparent strengthening of the ppt
condition, which will turn out to be useful in proving below that
a ppt state fulfills the Bell inequalities. Again,
the assumptions on $\cA$ and $\cB$ are of the generic type as
stated at the beginning of the section.
\begin{Lemma} \label{inequal}
Let $\o$ be a ppt state on $\cR$ for the bipartite system
$(\cA,\cB)\subset\cR$. Then for any choice of finitely many
$A_1,\ldots,A_k \in \cA$, and $B_1,\ldots,B_k \in \cB$, it holds
that
$$ |\o(T)|^2 \le\sum_{\a,\b} \o(A_{\b}A^*_{\a}B_{\a}^* B_{\b}) $$
where $T = \sum_{\a} A_{\a}B_{\a}$.
\end{Lemma}
{\it Proof.} We add new elements $A_0 = \eins$ and $B_0 = \lambda \eins$
for $\lambda \in \bC$ to the families $A_1,\ldots,A_k$,
$B_1,\ldots,B_k$. The condition of ppt then applies also with the new
families $A_0,A_1,\ldots,A_k \in \cA_1$, $B_0,B_1,\ldots,B_k \in
\cA_2$, entailing that
\begin{eqnarray*}
 0 \le \sum_{\a,\b = 0}^k \o(A_{\b}A^*_{\a}B_{\a}^* B_{\b}) & =&
 \sum_{\a,\b = 1}^k \o(A_{\b}A^*_{\a}B_{\a}^* B_{\b})\\ &+& \o(\lambda
 T^*) + \o(\overline{\lambda}T) + \o(|\lambda |^2 \eins)\,.
\end{eqnarray*}
Now insert $\lambda = -\o(T)$ and use that, since $\o$ is a state,
it holds that $\omega(T^*) = \overline{\o(T)}$. This yields
immediately the inequality claimed in Lemma~\ref{inequal}. \qed

In a similar spirit, we can apply the standard trick of {\it
polarization}, i.e., of replacing the arguments in a positive
definite quadratic form by linear combinations to get a condition
on a bilinear form. The {\it polarized version} of the
ppt-property is the following, and makes yet another connection to
the ordinary matrix version of the ppt-property:

\begin{Lemma}\label{polarize}
Let $\o$ be a state on a bipartite system $(\cA,\cB)\subset\cR$.
Then for any choice of elements $A_1,\ldots,A_n\in\cA$ and
$B_1,\ldots,B_m\in\cB$, introduce the $(nm)\times(nm)$-matrix $X$
by
\begin{equation}\label{gram}
    \langle i\alpha|X|j\beta\rangle
       =\o\bigl(A_iB_\alpha B_\beta^*A_j^*\bigr)\;.
\end{equation}
All such matrices are positive definite for any state $\o$.
Moreover, they all have a positive partial transpose if and only
if $\o$ is ppt.
\end{Lemma}

{\it Proof}\/: The positivity for arbitrary states says that, for
all complex $n\times m$-matrices $\Phi$, we have
\begin{equation}\label{posgram}
   \sum_{i\a j\b}\overline{\Phi_{i\a}}\,\langle
        i\alpha|X|j\beta\rangle\,\Phi_{j\b}
   =\o(X^*X)\geq0\;,
\end{equation}
where $X=\sum_{i\a}{\Phi_{i\a}}B_\a^*A_i^*$. For the ppt-property,
decompose an arbitrary $\Phi$ as
$$\Phi_{i\a}=\sum_\mu u_{i\mu}v_{\a\mu}\;,$$
 for suitable coefficient matrices $u,v$. For example, we can get
$u$ and $v$ from the singular value decomposition of $\Phi$.
Inserting this into the condition for the positivity of $X^{T_2}$,
we find
\begin{eqnarray}
     \sum_{i\a j\b}\overline{\Phi_{i\a}}\,
        \langle i\alpha|X^{T_2}|j\beta\rangle\,\Phi_{j\b}
      &=&\sum_{i\a j\b}\overline{\Phi_{i\a}}\,
        \langle i\b|X|j\a\rangle\,\Phi_{j\b}
          \nonumber\\
      &=&\sum_{i\a j\b\mu\nu}
          \overline{u_{i\mu}}\;\overline{v_{\a\mu}}\,
          u_{j\nu}\,v_{\b\nu}\;
          \o(A_iB_\b B^*_\a A^*_j)
          \nonumber\\
      &=&\sum_{\mu\nu}\o(\widetilde A_\mu\widetilde A^*_\nu
                        \widetilde B^*_\nu\widetilde B_\mu)\;,
      \nonumber
\end{eqnarray}
with
$$ \widetilde A_\mu=\sum_i\overline{u_{i\mu}}A_i
    \quad\mbox{and}\quad
   \widetilde B_\mu=\sum_\a\overline{v_{\a\mu}}B^*_\a  \;.
$$
The ppt-property demands that all these expressions are positive,
and conversely, positivity of all these expressions entails that
$\omega$ is ppt. \qed

This Lemma greatly helps to sort the big mess of indices which
would otherwise clutter the proof of the following result. It
contains as a special case the observation that the tensor product
of ppt states is ppt, provided we consistently maintain the
Alice/Bob distinction, which will be important for establishing
the preservation of the ppt-property under general distillation
protocols. In the standard case this is an easy property of the
partial transposition operation. Since this is not available in
general, we have to give a separate proof based on our definition.

\begin{Lemma} \label{lem:tensor}
Let $(\cA_k,\cB_k)\subset\cR_k$ be a finite collection of
bipartite systems, all contained in a common ambient algebra $\cR$
such that all algebras $\cR_k$ commute. Let $\cA$ (resp.\ $\cB$)
denote the C*-algebra generated by all the $\cA_k$ (resp.\
$\cB_k$). Let $\o$ be a state on $\cR$, which is ppt for each
subsystem, and which factorizes over the different $\cR_k$. Then
$\o$ is ppt for $(\cA,\cB)\subset\cR$.
\end{Lemma}

{\it Proof}\/. We show the ppt property in polarized form. Since
$\cA$ is generated by the commuting algebras $\cA_k$, we can
approximate each element by linear combinations of products
$A=\prod_kA^{(k)}$. Since the polarized ppt-condition is
continuous and linear in $A_i$, it suffices to prove it for
choices $A_i=\prod_kA^{(k)}_i$, and similarly for $B_\a$. For such
choices the factorization of $\o$ implies that the $X$-matrix from
the Lemma is the tensor product of the matrices $X_k$ obtained for
the subsystems. The partial transposition of the whole matrix is
done factor by factor, and since all the $X_k^{T_2}$ are positive,
so is their tensor product $X^{T_2}$. \qed

We close this section by pointing out a mathematically more
elegant way of expressing the ppt property. It employs the concept
of the {\it opposite algebra} $\cA^{\rm op}$ of a given
$*$-algebra $\cA$. The opposite algebra is the $*$-algebra formed
by $\cA$ with its original vector addition, scalar multiplication,
and adjoint (and operator norm), but endowed with a new algebra
product:
$$ A \bullet B = BA\,, \quad A,B \in \cA\;,$$
where on the right hand side we read the original algebra product
of $\cA$. There is a linear, $*$-preserving, one-to-one, onto map
$\theta: \cA^{\rm op} \to \cA$ given by $\theta(A) = A$, which is
an anti-homomorphism (i.e., $\theta(A\bullet B) =
\theta(B)\theta(A)$ for all $A,B \in \cA^{\rm op}$. With its help
one can define a linear, $*$-preserving map $\theta\odot{\rm
id}:\cA^{\rm op} \odot \cB\to \cR$ by
$$(\theta \odot{\rm id})(A \odot B) = \theta(A)B\;,$$
where we have distinguished the ``algebraic tensor product''\
$\odot$, i.e., the tensor product as defined in linear algebra,
from the ordinary tensor product ``$\otimes$'' of C*-algebras,
which also contains norm limits of elements in $\cA\odot\cB$.

By definition, $(\theta \odot{\rm id})$ has dense range, but is
usually unbounded, and does not preserve positivity. Given any
state $\omega$ on $\cR$, it induces a linear functional
$\omega_{\theta \odot{\rm id}} = \omega \circ (\theta\odot{\rm
id})$ on $\cA^{\rm op} \odot \cB$. Then it is not difficult to
check that the functional $\omega_{\theta \odot{\rm id}}$ is
positive (i.e.\ $\omega_{\theta \odot{\rm id}}(C^*C) \ge 0$ for
all $C \in\cA^{\rm op} \odot \cB$) if and only if $\omega$ is a ppt
state.

It would be interesting to study ``mild failures'' of the ppt
condition, i.e., cases in which $\omega_{\theta \odot{\rm id}}$,
although not positive, is a bounded linear functional, or maybe
even a normal linear functional on $\cA^{\rm op} \odot \cB$.

\section{Relation to the Bell-CHSH Inequalities}
\setcounter{equation}{0}
Now we study the connection of the ppt-property to
Bell-inequalities in the CHSH form \cite{CHSH}. Again, we have to
recall some terminology. A state $\o$ on a bipartite system
$(\cA,\cB)\subset\cR$ is said to {\it satisfy the Bell-CHSH inequalities
if}
\begin{equation}
\label{BellIn}
|\o(A(B' + B) + A'(B' -B))| \le 2
\end{equation}
holds for all hermitean $A,A' \in \cA$ and $B,B' \in \cB$ whose
operator norm is bounded by 1. A quantitative measure of the
failure of a state to satisfy the Bell-CHSH inequalities is measured by
the quantity
$$ \beta(\o)=\sup_{A,A',B,B'}\,\o(A(B' + B) + A'(B' -B)) \;.$$
where the supremum is taken over all  admissible $A,A',B,B'$ as in
(\ref{BellIn}). By Cirel'son's inequality \cite{Cirelson},
\begin{equation}\label{cirelson}
  \beta(\o)\leq2\sqrt2\;.
\end{equation}
If equality holds here, we say that the bipartite state $\o$ {\it
violates the Bell-CHSH inequalities maximally}. The proof of the
following result is adapted from the finite dimensional case
\cite{WWolf}.
\begin{Theorem} \label{Bell}
If a bipartite state is ppt, then it satisfies the Bell-CHSH
inequalities.
\end{Theorem}
{\it Proof. } The right hand side in \eqref{BellIn} is linear in
each of the arguments $A,A' \in \cA$ and $B,B' \in \cB$. Hence we
can search for the maximum of this expression taking each of these
four variables as an extreme point of the admissible convex
domain. The extreme points of the set hermitean $X$ with
$||X||\leq1$ are those with $X^2=\idty$. Hence it is sufficient to
show that the bound \eqref{BellIn} holds for all hermitean
arguments fulfilling $A^2 = A'{}^2 = B^2 = B'{}^2 = \eins$. For
such operators $A,A'$ and $B,B'$ we set, following \cite{Lan1},
$$ C = A(B' + B) + A'(B' -B) $$
and obtain
\begin{equation}
\label{u}
|\o(C)|^2 \le \o(C^2) = 4 + \o([A,A'][B,B'])
\end{equation}
where $[X,Y] = XY - YX$ denotes the commutator. On the other hand, if
we set $A_1 = A$, $A_2 = A'$, $B_1 = B' + B$, $B_2 = B' -B$, we get
according to Lemma \ref{inequal}, since $\o$ admits a ppt,
\begin{eqnarray}
\label{v}
|\o(C)|^2 & \le & \sum_{\a,\b = 1}^2 \o(A_{\b}A^*_{\a}B_{\a}^*B_{\b})
\nonumber \\
 & = & 4 - \o([A,A'][B,B'])\,.
\end{eqnarray}
Adding \eqref{u} and \eqref{v} yields $|\o(C)|^2 \le 4$ which is
equivalent to \eqref{BellIn}. \qed

\section{Distillability for General Quantum Systems}
\setcounter{equation}{0}
If entanglement is considered as a resource provided by some
source of bipartite systems, it is natural to ask whether the
particular states provided by the source can be used to achieve
some tasks of quantum information processing, such as
teleportation. Usually the pair systems provided by the source are
not directly usable, so some form of preprocessing may be
required. This upgrading of entanglement resources is known as
distillation. The general picture here is that the source can be
used several times, say $N$ times. The allowed processing steps
are local quantum operations, augmented by classical communication
between the two labs holding the subsystems (``LOCC operations''
\cite{BDSW,Keyl}, see also \cite{?c}), usually personified by the two physicists
operating the labs, called Alice and Bob. That is, the decision
which operation is applied by Bob can be based on measuring
results previously obtained by Alice and conversely. The aim is to
obtain, after several rounds of operations, some bipartite quantum
systems in a state which is nearly maximally entangled. The number
of these systems may be much lower than $N$, whence the name
``distillation''.

The idea of distillation can be generalized to combinations of
 resources. For example, a bound entangled (i.e., not by
 itself  distillable) state can sometimes be utilized to improve
 entanglement in another state \cite{activate}.

The optimal {\it rate} of output particles per input particle is
an important quantitative measure of entanglement in the state
produced by the source. Distillation rates are very hard to
compute because they involve an optimization over all distillation
procedures, a set which is difficult to parameterize. A simpler
question is to decide whether the rate is zero or positive. In the
latter case the state is called {\it distillable}.

 In this paper
we will look at two types of results on distillability,
ensuring either success or failure: 
 We will show that many states in quantum field
theory are distillable, by using an especially simple kind of
distillation protocol. States for which this works are also called
{\it 1-distillable} (see below).

On the other hand we will show that distillable states cannot be
ppt. Note that this is a statement about all possible LOCC
protocols, so we will need to define this class of operations more
precisely in our general context. The desired implication will
become stronger if we allow more operations as LOCC, so we should
make only minimal technical assumptions about this class of
operations. To begin with, LOCC operations are operations between
different bipartite systems. So let $(\cA_1,\cB_1)\subset\cR_1$
and $(\cA_2,\cB_2)\subset\cR_2$ be bipartite systems. An operation
localized on the Alice side will be a completely positive map
$T:\cA_1\to\cA_2$ with $T(\eins)=\eins$. Note that since we
defined the operation in terms of observables, we are working in
the Heisenberg picture, hence $1$ labels the output system and $2$
labels the input system. An operation also producing classical
results is called an {\it instrument} in the terminology of Davies
\cite{Davies}. When there are only finitely many possible
classical results, this is given by a collection $T_x$ of
completely positive maps, labelled by the classical result $x$,
such that $\sum_xT_x(\eins)=\eins$. Similarly, an operation
depending on a classical input $x$ is given by a collection of
completely positive maps $S_x$ such that $S_x(\eins)=\eins$.
Hence, whether the classical parameter $x$ is an input or an
output is reflected only in the normalization conditions. A LOCC
operation with information flow only from Alice to Bob is then
given by a completely positive map $M:\cR_1\to\cR_2$ such that
\begin{equation}\label{e:locc}
    M(AB)=\sum_xT_x(A)S_x(B)\;,
\end{equation}
where the sum is finite, and for each $x$, $T_x:\cA_1\to\cA_2$ and
$S_x:\cB_1\to\cB_2$ are completely positive with the normalization
conditions specified above. This will be the first round of a LOCC
protocol. In the next round, the flow of information is usually
reversed, and all operations are allowed to depend on the
classical parameter $x$ measured in the first round. Iterating
this will lead to a similar expression as (\ref{e:locc}), with $x$
replaced by the accumulated classical information obtained in all
rounds together. The normalization conditions will depend in a
rather complicated way on the information parameters of each
round. However, as is easily seen by induction the overall
normalization condition
\begin{equation}\label{sepsupop}
   \sum_xT_x(\eins)S_x(\eins) =\eins
\end{equation}
will also hold for the compound operation. Fortunately, we only
need this simple condition. An operator $M$ of the form
(\ref{e:locc}), with completely positive $T_x,S_x$, but with only
the overall normalization condition (\ref{sepsupop}), is called a
{\it separable superoperator}, in analogy to the definition of
separable states. More generally, we use this term also for limits
of such operators $M_\alpha$, such that probabilities converge for
all input states, and all output observables. By such limits we
automatically also cover the case of continuous classical
information parameters $x$, in which the sums are replaced by
appropriate integrals.

Then we can state the following implication:

\begin{Proposition} Let $M$ be a separable superoperator between
bipartite systems $(\cA_i,\cB_i)\subset\cR_i$, ($i=1,2$), and let
$\o_2$ be ppt. Then the output state $\o_1(X)=\o_2(M(X))$ is also
ppt. In particular, ppt states are not distillable with LOCC
operations.
\end{Proposition}
{\it Proof. } The ppt-preserving property is preserved under
limits as described above, and also under sums, so it suffices to
consider superoperators $M$, in which the sum (\ref{e:locc}) has
only a single term, i.e., $\o_1(AB)=\o_2(T(A)S(B))$.

 Let
 $A_1,\ldots,A_k \in \cA_1$, and $B_1,\ldots,B_k \in \cB_1$.
We have to show that\\
 $\sum_{\a,\b}\o_2\left( T(A_{\b}A^*_{\a})S(B_{\a}^*
                                            B_{\b})\right)\geq0$.
Now because $T$ is completely positive, the matrix
$T(A_{\b}A^*_{\a})$ is positive in the algebra of $\cA_2$-valued
$k\times k$-matrices, and hence we can find elements
$t_{n\a}\in\cA_2$ (the matrix elements of the square root) such
that
\begin{equation}\label{tdecA}
  T(A_{\b}A^*_{\a})=\sum_n(t_{n\b})^*t_{n\a}\;.
\end{equation}
Of course, there is an analogous decomposition
\begin{equation}\label{tdecB}
  S(B_{\b}B^*_{\a})=\sum_m(s_{m\b})^*s_{m\a}\;.
\end{equation}
Hence, observing the changed order of the indices $\a,\b$ in the
$S$-term:
\begin{equation}
\sum_{\a,\b}\o_1\left( T(A_{\b}A^*_{\a})S(B_{\a}^*
                                            B_{\b})\right)
   =\sum_{n,m}\sum_{\a,\b}\o_2\left((t_{n\b})^*t_{n\a}\
                                      (s_{m\a})^*s_{m\b}\right)
   \;,
   \nonumber
\end{equation}
which is positive, because the input state $\o_2$ was assumed to
be ppt.

For distillability we have to consider tensor powers of the given
state and try to obtain a good approximation of a singlet state of
two qubits by some LOCC
 operation. However, since the final state
is clearly not ppt, and the input tensor power is ppt by
Lemma~\ref{lem:tensor}, the statement just proved shows that this
impossible. \qed

For positive distillability results it is helpful to reduce the
vast complexity of all LOCC operations, applied to arbitrary
tensor powers, and to look for specific simple protocols for the
case at hand.  Since we are not concerned with rates, but only
with the yes/no question of distillability, some major
simplifications are possible. The first simplification is to
restrict the kind of classical communication. Suppose that the
local operations are such that every time they also produce a
classical signal ``operation successful'' or ``operation failed''.
Then we can agree to use only those pairs in which the operation
was successful on both sides. In all other cases we just try
again. Note that this requires two-way classical communication,
since Alice and Bob both have to give their ok for including a
particular pair in the ensemble. However, in the simplest case no
further communication between Alice and Bob is used. To state this
slightly more formally, let $T$ denote the distillation operation
in such a step, written in the Heisenberg picture. This is a
selective operation in the sense that $T(\idty)\leq\idty$, and
$\omega\bigl(T(\idty)\bigr)$ is the probability for successfully
obtaining a pair. Then by the law of large numbers we can build
from this a sequence of non-selective distillation operations on
many such pairs, which produce systems in the state
\begin{equation}\label{renorm}
    \omega^{[T]}(A)=\frac{\omega\bigl(T(A)\bigr)}{\omega\bigl(T(\idty)\bigr)}
\end{equation}
with rate close to the probability $\omega\bigl(T(\idty)\bigr)$.
If we are only interested in the yes/no question of distillability
and not in the rate, then obviously selective operations are just
as good as non-selective ones. Moreover, it is sufficient for
distillability that $\omega^{[T]}$ be distillable for some such
$T$. It is also convenient to restrict the type of output systems:
it suffices to produce a pair of qubits ($2$-level systems) in a
distillable state, because from a sufficient number of such pairs
any entangled state can be generated by LOCC operations. Any
target state which has non-positive partial transpose will do,
because for qubits ppt and non-distillability are equivalent.
Finally, we look at situations where the criterion can be applied
without going to higher tensor powers. In the simplest case only one
pair prepared in the original state $\omega$ is needed to obtain a
distillable qubit pair with positive probability.

\begin{Definition}\label{def:1dist}
 A state $\o$ on a bipartite system
$(\cA,\cB)\subset\cR$ is called {\bf 1-distillable}, if there are
completely positive maps $T:\Bofcc\to\cA$ and $S:\Bofcc\to\cB$
such that the functional $\o_2(X\otimes Y)=\o(T(X)S(Y))$,
$X\otimes Y \in B(\bC^2 \otimes \bC^2)$, on the
two-qubit system is not ppt.
\end{Definition}

Then according to the discussion just given, 1-distillable states
are distillable. If the maps $T,S$ are normalized such that
$||T(\eins)||=||S(\eins)||=1$, and $\o_2$ is close to a multiple
of a singlet state, a rough estimate of the distillation rate
achievable from $\o$ is the normalization constant
$\o_2(\eins)$. In the field theoretical applications below this
rate will be very small.

Note that specifying a completely positive map $T:\Bofcc\to\cA$ is
equivalent to specifying the four elements
$T_{k\ell}=T(|k\rangle\langle\ell|)\in\cA$ or, in other words, an
$\cA$-valued $2\times2$-matrix, called the {\it Choi matrix} of
$T$. It turns out that $T$ is completely positive iff the Choi
matrix is positive in the algebra of such matrices (isomorphic to
$\cA\otimes\Bofcc$). This allows a partial converse of the
implication ``distillable $\Rightarrow$ npt'':

\begin{Lemma}Let $\o$ be a state on a bipartite system
$(\cA,\cB)\subset\cR$, and suppose that the ppt condition in
Definition~\ref{def:ppt} fails already for $k=2$. Then $\o$ is
1-distillable in the sense of Definition~\ref{def:1dist}.
\end{Lemma}

{\it Proof\/}: Let $A_1,A_2,B_1,B_2$ be as in
Definition~\ref{def:ppt}. Then we can take the matrix
 $A_\a A_\b^*$ as the Choi matrix of $T$, i.e., with a similar
 definition for $S$:
\begin{eqnarray}\label{choi22}
    T(M)&=&\sum_{\a,\b} A_\b \langle\b|M|\a\rangle A_\a^*
              \nonumber\\
    S(N)&=&\sum_{\a',\b'} B_{\a'}^* \langle\a'|N|\b'\rangle B_{\b'}
            \;.  \nonumber
\end{eqnarray}
Inserting this into Definition~\ref{def:1dist}, we find
\begin{equation}\label{choi22x}
    \o_2(Z)=\sum_{\a,\b,\a'\b'}\langle\b\a'|Z|\a\b'\rangle
              \o\Bigr(A_\b A_\a^*B_{\a'}^*B_{\b'}\Bigl)\,, \quad Z \in B(\bC^2 \otimes \bC^2)\,.
\end{equation}
In particular, when $Z$ is equal to the transposition operator
$Z\vert\a\b'\rangle=\vert\b'\a\rangle$, this expectation is equal
to the sum in Definition~\ref{def:ppt}, hence negative by
assumption. On the other hand, $Z$ has a positive partial
transpose (proportional to the projection onto a maximally
entangled vector), hence $\o_2$ cannot be positive. \qed

\section{The Reeh-Schlieder property}
\setcounter{equation}{0}
In this section we will establish a criterion for
1-distillability which will be useful in quantum field
theoretical applications. We prove it in an abstract form, which
for the time being makes no use of spacetime structure. We will
assume that all observable algebras are given as operator
algebras, i.e., we look at bipartite systems of the kind
$(\cA,\cB)\subset\Bof\cH$. This is no restriction of generality,
since every C*-algebra (here the ambient algebra of the bipartite
system) may be isomorphically realized as an algebra of operators.
The non-trivial information contained in any such representation
is about a special class of states, namely the normal ones (see
\eqref{e:normal}). Any state of a C*-algebra becomes normal in a
suitable representation, so the choice of representation is mainly
the choice of a class of states of interest. In particular, we
have the  {\it vector states} on $\Bof\cH$, which are states of
the form
\begin{equation}\label{vecstates}
    \o_{\psi}(R) = \langle \psi | R | \psi \rangle \,, \quad R \in \Bof\cH\,,
\end{equation}
with $\psi\in\cH$ a unit vector. Again, this is not a loss of
generality, since every bipartite system can be written in this
way, by forming the GNS-representation \cite{BratRob1} of the
ambient algebra\footnote{For a state $\omega$ on a $C^*$-algebra $\cR$,
there is always a triple $(\pi_\omega,\cH_\omega,\Omega_\omega)$ where:
(1) $\pi_\omega$ is a $*$-preserving representation of $\cR$ by bounded
linear operators on the Hilbert-space $\cH_\omega$. (2) $\Omega_\omega$
is a unit vector in $\cH_\omega$ so that $\pi_\omega(\cR)\Omega_\omega$ is dense
in $\cH_\omega$. (3) $\omega(R) = \langle \Omega_\omega | \pi_\omega(R) | \Omega_\omega \rangle$
for all $R \in \cR$. $(\pi_\omega,\cH_\omega,\Omega_\omega)$ is called the
GNS representation of $\omega$; see, e.g., \cite{BratRob1} for its construction.}.
 However, in this language the key condition of
this section is more easily stated. It has two formulations: one
emphasizing the operational content from the physical point of
view, and one which is somewhat simpler mathematically. We state
their equivalence in the following Lemma (whose proof is entirely
trivial).

\begin{Lemma} Let $\cA\subset\Bof\cH$ be a C*-algebra, and
$\psi\in\cH$ a unit vector. Then the following are equivalent:\\
(1) $\psi$ has the {\bf Reeh-Schlieder property} with respect to
$\cA$, i.e., for each unit vector $\chi \in \cH$ and each
$\varepsilon > 0$, there is some $A\in \cA$, so that
$$ |\,\o_{\chi}(R) - \o_{\psi}(A^*RA)/\o_{\psi}(A^*A)\,| < \varepsilon ||R||$$
holds for all $R \in \Bof\cH$.\\
(2) $\psi$ is {\bf cyclic} for $\cA$, i.e., the set
 $\cA \psi = \{A\psi: A \in \cA\}$ is dense in $\cH$.
\end{Lemma}
We also remark that a vector $\psi$ in $\cH$ is called {\it separating}
for $\cA$ if for each $A \in \cA$, the relation $A\psi = 0$ implies
that $A = 0$. It is a standard result in the theory of operator algebras
that $\psi$ is cyclic for a von Neumann algebra $\cA$ if and only
if $\psi$ is separating for its commutant $\cA'$ (see, e.g., \cite{BratRob1}).
Note that $\cA$, a subset of $\Bof\cH$, is a von Neumann algebra if it
coincides with its bicommutant $\cA''$, where for $\cB \subset \Bof\cH$,
its commutant is the von Neumann algebra $\cB' =\{ R \in \Bof\cH: RB = BR\ \forall
B \in \cB\}$.

The physical meaning of the Reeh-Schlieder property is that any
vector state on $\Bof\cH$ can be obtained from $\omega_{\psi}$ by selecting according
to the results of a measurement on the subsystem $\cA$. Let us
denote by $A_1$ a multiple of the $A$ from the Lemma, normalized
so that $||A_1||\leq1$, and set $A_0=(\eins-A_1^*A_1)^{1/2}$. Then
the operation elements $T_i(R)=A_i^*RA_i$ ($i=0,1$) together
define an instrument. The operation without selecting according to
results is $T(R)=T_0(R)+T_1(R)$. This instrument is {\it
localized} in $\cA$ in the sense that $T_i(\cA)\subset\cA$, and
that for any $B$ commuting with $\cA$, in particular for all
observables of the second subsystem of a bipartite system, we get
$T(B)=B$. That is, no effect of the operation is felt for
observables outside the subsystem $\cA$. Of course, $T_i(B)\neq
B$, but this only expresses the state change by selection in the
presence of correlations. The state appearing in the Reeh-Schlieder
property is just a selected state, obtained by running the
instrument on systems prepared according to $\o_\psi$, and keeping
only the systems with a 1-response. By taking convex combinations
of operations, one can easily see that also every convex
combination of vector states, and hence any normal state can be
approximately obtained from $\o_\psi$.

Our next result connects these properties with distillability.
\begin{Theorem} \label{RS-S-D}
Let $(\cA,\cB)\subset\Bof\cH$ be a bipartite system, with both
$\cA,\cB$ non-abelian. Suppose $\psi\in\cH$ is a unit  vector
which has the Reeh-Schlieder property with respect to $\cA$. Then
$\o_\psi$ is 1-distillable.
\end{Theorem}
The proof of this statement takes up ideas of Landau, and utilizes
Lemma 5.5 in \cite{SJS-RMP2}. To keep our paper self-contained, we
nevertheless give a full proof here.

{\it Proof, step 1}\/:  We first treat the special case in which
$\cA$ and $\cB$ are von Neumann algebras, i.e., of algebras also
closed in the weak operator topology. Then a theorem  due to
M.~Takesaki \cite{Tak} asserts that there are non-vanishing
*-homomorphisms $\tau:\Bofcc\to\cA$ and $\sigma:\Bofcc\to\cB$,
which may, however, fail to preserve the identity. Consider the
map $\pi:\Bofcc\otimes\Bofcc\to\Bof\cH$, given by
\begin{equation}\label{togetherep}
    \pi(X\otimes Y)=\tau(X)\sigma(Y)\;.
\end{equation}
One easily checks that, because the ranges of $\tau$ and $\sigma$
commute, $\pi$ is a *-homo\-mor\-phism. But as a C*-algebra
$\Bofcc\otimes\Bofcc\cong\Bof{\bC^4}$ is a full matrix algebra.
Since this has no ideals, $\pi$ is either an isomorphism or zero.

{\it Step 2}\/: We have to show that $\tau$ can be chosen so that
$\pi$ is non-zero. In many situations of interest this would
follow automatically because both $\tau(\eins)$ and
$\sigma(\eins)$ are non-zero: Often $\cA$ and $\cB$ also have the
so-called Schlieder property \cite{Schl} (an independence property
\cite{SJS-RMP2}), which means that $A\in\cA$, $B\in\cB$,
$A,B\neq0$ imply $AB\neq0$. (There seems to be an oversight in
\cite[Lemma 5.5]{SJS-RMP2} concerning this assumption.) However,
we do not assume this property, and instead rely once again on the
Reeh-Schlieder property of $\cA$.

Let us take $\sigma$ as guaranteed by Takesaki's Theorem, and set
$p=\sigma(\idty)$. Then if $p\cA p$ is non-abelian, we can apply
Takesaki's result to this algebra, and find a homomorphism $\tau$
with $\pi(\eins)=\tau(\eins)p=\tau(\eins)\neq0$. So we only need
to exclude the possibility that $p\cA p$ is abelian.

In other words, we have to exclude the possibility that in some
Hilbert space $\cH_{(p)}\equiv p\,\cH$ there is an abelian von
Neumann algebra $\cA_{(p)}\equiv p\cA p$ with a cyclic vector
$\psi_{(p)}=p\psi$, so that $\cA_{(p)}$ commutes with a
copy $\cB_{(p)}\equiv\sigma(\Bofcc)$ of the
$2\times2$-matrices. The latter property entails that $({\cA}_{(p)})'$
is non-abelian.

We will exclude this possibility by adopting it as a hypothesis and
showing that this leads to a contradiction. Let $q$ denote
the projection onto the subspace of $\cH_{(p)}$ generated by
$({\cA}_{(p)})'{\psi}_{(p)}$. This projection is contained in
$({\cA}_{(p)})'' = {\cA}_{(p)}$. Let ${\cH}_{(qp)} =
q\cH_{(p)}$, then $\psi_{(qp)} = q\psi_{(p)} = \psi_{(p)} \in {\cH}_{(qp)}$
is both a cyclic and separating vector for the von Neumann algebra
$(\cA_{(p)})'_{(q)} = q({\cA}_{(p)})'q$,
and since ${\psi}_{(p)}$ is separating for $({\cA}_{(p)})'$
(owing to the assumed cyclicity of $\psi_{(p)}$ for ${\cA}_{(p)}$),
$({\cA}_{(p)})'_{(q)}$ is non-abelian since so is
$(\cA_{(p)})'$ by hypothesis. On the other hand, abelianess of
${\cA}_{(p)}$ entails that $ \cA_{(qp)} = ({\cA}_{(p)})_{(q)} = 
q({\cA}_{(p)})q$ is a von Neumann algebra in $B({\cH}_{(qp)})$
for which $({\cA}_{(p)})'_{(q)} = (\cA_{(qp)})'$,
where the second commutant is taken in $B(\cH_{(qp)})$. Clearly,
$\cA_{(qp)}$ is again abelian. However, since ${\psi}_{(qp)}$
is cyclic and separating for 
$(\cA_{(p)})'_{(q)} = (\cA_{(qp)})'$,
it follows by the Tomita-Takesaki theorem \cite{BratRob1} that $\cA_{(qp)}$
is anti-linearly isomorphic to $(\cA_{(qp)})'$, which is a
contradiction in view of the abelianess of $\cA_{(qp)}$ and
non-abelianess of $(\cA_{(qp)})'$. 

 To summarize, we have shown that with a suitable
$\tau$, the representation $\pi$ in \eqref{togetherep} is an
isomorphism.

{\it Step 3}\/: Now consider the singlet vector
$\Omega=(|+-\rangle-|-+\rangle)/\sqrt2\in(\bC^2\otimes\bC^2)$.
Since $\pi$ has trivial kernel, the projection
$Q=\pi(|\Omega\rangle\langle\Omega|)$ is non-zero, and hence there
is a vector $\chi$ in the range of this projection. Obviously,
$\o_\chi(\pi(Z))
   =\o_\chi(Q\pi(Z)Q)
   =\o_\chi(\pi(|\Omega\rangle\langle\Omega|Z|\Omega\rangle\langle\Omega|))
   =\langle\Omega|Z|\Omega\rangle\o_\chi(Q)
   =\langle\Omega|Z|\Omega\rangle$ holds for all $Z \in B(\bC^2 \otimes \bC^2)$.
Now we introduce the distillation maps $T,S$ of
Definition~\ref{def:1dist}. On Bob's side $S(Y)=\sigma(Y)$ is good
enough. For Alice we take $T(X)=A^*\tau(X)A$, where $A\in\cA$ is
the operator from the Reeh-Schlieder property for some small
$\varepsilon>0$. The functional distilled from this is
\begin{eqnarray}\label{om2dist}
 \o_2(X\otimes Y)
     &=&\o_\psi(T(X)S(Y))
      =\o_\psi(A^*\tau(X)A \sigma(Y))
        \nonumber\\
     &=&\o_\psi(A^*\tau(X)\sigma(Y)A )
      =\o_\psi(A^*\pi(X\otimes Y)A )\;.
       \nonumber
\end{eqnarray}
Now the Reeh-Schlieder property, applied to the operator
$R=\pi(Z)\in\Bof\cH$, asserts that $\o_2(Z)/\o_2(\eins)$ is close
to $\o_\chi(\pi(Z))=\langle\Omega|Z|\Omega\rangle$, $Z \in
 B(\bC^2 \otimes \bC^2)$  . Hence, up to
normalization, $\o_2$ is close to a singlet state, and therefore
is not ppt. This proves the Theorem in the case that $\cA$ and
$\cB$ are von Neumann algebras.

{\it Step 4}\/: When the C*-algebras $\cA,\cB$ satisfy the
assumptions of the Theorem, so do their weak closures, the von
Neumann  $\cA'',\cB''$: since $\cA\subset\cA''$ these algebras are
both non-abelian, and by taking commutants of the inclusion
$\cB\subset\cA'$, we get the commutation property
$\cA''\subset\cB'$ of the von Neumann algebras. Of course, if
$\cA\psi$ is dense in $\cH$, so is the larger set $\cA''\psi$.

Now let $T'':\Bofcc\to\cA''$ and $S'':\Bofcc\to\cB''$ be the
distillation maps, whose existence we have just proved. We have to
find maps $T:\Bofcc\to\cA$, and $S:\Bofcc\to\cB$ with smaller
ranges, which do nearly as well. This is the content of the
following

\begin{Lemma} Let $\cA\subset\Bof\cH$ be a C*-algebra, and let
$k\in\bN$.  Consider a completely positive map
$T:\Bof{\bC^k}\to\cA''$. Then for any finite collection of vectors
$\phi_1,\ldots,\phi_n$, and $\varepsilon>0$ we can find a
completely positive map $\widetilde T:\Bof{\bC^k}\to\cA$ such
that, for all $X\in\Bof{\bC^k}$, and all $j$, we have
$||(T(X)-\widetilde T(X))\phi_j||\leq\varepsilon||X||$.
\end{Lemma}

Obviously, with such approximations (for just the single vector
$\phi_1=\psi$), we get a distilled state $\o_2$ arbitrarily close
to what we could get from the distillation in the von Neumann
algebra setting. This concludes the proof of the Theorem, apart
from the proof of the Lemma.

{\it Proof of the Lemma}\/: Note that the version of the Lemma
with $k=1$ just states that the positive cone of $\cA$ is strongly
dense in the positive cone of $\cA''$, which is a direct
consequence of Kaplansky's Density Theorem \cite[Theorem
4.8]{Tak}. We will reduce the general case to this by
parameterizing all completely positive maps
$T_i:\Bof{\bC^k}\to\cA''$ by their Choi-matrices
\begin{equation}\label{choikap}
    t_i=\sum_{\a\b=1}^k T_i(|\a\rangle\langle\b|)
        \otimes|\a\rangle\langle\b|  \in\cA''\otimes\Bof{\bC^k} \;,
\end{equation}
where ``subscript $i$'' equals ``tilde'' or ``no tilde''. Note
that $\cA''\otimes\Bof{\bC^k}$ is the von Neumann algebra closure
of $\cA\otimes\Bof{\bC^k}$, so via Kaplansky's Density Theorem we
obtain, for the given positive element
$t\in\cA''\otimes\Bof{\bC^k}$, and any finite collection of
vectors in $\cH\otimes\bC^k$, a positive approximant $\tilde
t\in\cA\otimes\Bof{\bC^k}$. As the collection vectors we take the
given $\phi_i$, tensored with the basis vectors of $\bC^k$, which
implies the desired approximation for all $X$, which are matrix
units $|\a\rangle\langle\b|$. However, because $k$ is finite, and
all norms are equivalent on a finite dimensional vector space, we
can achieve a bound as required in the Lemma.
 \qed

We remarked in the beginning of this section that assuming the
given bipartite state to be a vector state in some representation
is not a restriction of generality. Therefore there should be a
version of the Theorem, which does not require a representation.
Indeed, we can go to the GNS-representation of the algebra
generated by $\cA$ and $\cB$ in the given state, and just restate
the conditions of the Theorem as statements about expectations in
the given state. This leads to the following

\begin{Corollary} Let $\o$ be a state on a bipartite system
$(\cA,\cB)\subset\cR$, and suppose that
 \begin{enumerate}
 \item For some $A\in\cA$, and $B_1,\ldots,B_4\in\cB$,
    $\o\bigl(AB_1[B_2,B_3]B_4\bigr)\neq0$,
    and a similar condition holds with $\cA$ and $\cB$
    interchanged.
 \item For all $B\in\cB$ and $\varepsilon>0$ there is an $A\in\cA$
 such that
    $$\o\Bigl((A-B)^*(A-B)\Bigr)\leq\varepsilon\;.$$
 \end{enumerate}
 Then $\o$ is 1-distillable.
\end{Corollary}

This opens an interesting connection with the theory of {\it
maximally entangled states} on bipartite systems. These are
generalizations of the EPR state, and have the property that for
every (projection valued) measurement on Alice's side there is a
``double'' on Bob's side such that if the two are measured
together the results agree with probability one \cite{KSW}. The
equation which has to be satisfied by Alice's observable $A$, and
the double $B$ looks very much like condition 2 in the Corollary
with $\varepsilon=0$, except that in addition one requires
$\o\Bigl((A-B)(A-B)^*\Bigr)=0$.

Before going to the context of quantum field theory, let us
summarize the implications we have established for a state $\o$ on
a general bipartite system $(\cA,\cB)\subset\cR$:

 \vskip6pt
\begin{center}
\epsfig{file=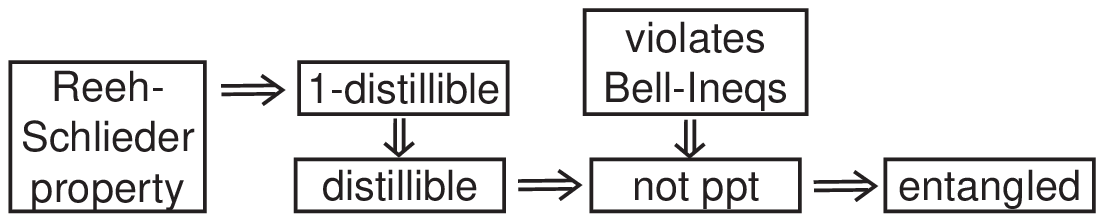,width=9cm}
\end{center}

{\small {\bf Figure 1}. Implications valid for any bipartite
state.}

\section{Distillability in Quantum Field Theory}\label{sec:qft1}

The generic occurrence of distillable states in quantum field theory
can by Theorem \ref{RS-S-D} be deduced from the fact that the Reeh-Schlieder
property, and non-abelianess, are generic features of von Neumann algebras
$\cA$ and $\cB$ describing observables localized in spacelike
disjoint regions $O_{\rm A}$ and $O_{\rm B}$ in relativistic
quantum field theory. To see this
more precisely, we have to provide a brief description of the basic
elements of quantum field theory in the operator algebraic
framework. The reader is referred to the book by R.\ Haag \cite{Haag}
for more details and discussion.

The starting point in the operator algebraic approach to quantum
field theory is that each system is described in terms of a
so-called ``net of local observable algebras'' $\{\fA(O)\}_{O
\subset \bR^4}$. This is a family of $C^*$-algebras indexed by the
open, bounded regions $O$ in $\bR^4$, the latter being identified
with Minkowski spacetime.  In
other words, to each open bounded region $O$ in Minkowski
spacetime one assigns a $C^*$-algebra $\fA(O)$, and it is required
that the following assumptions hold:
\begin{itemize}
\item[(I)] Isotony: $O_1 \subset O_2 \ \Rightarrow\ \fA(O_2) \subset
\fA(O_2)$\,,
\item[(II)] Locality: If the region $O$ is spacelike to the
region $O'$, then $AA' = A'A$ for all $A \in \fA(O)$ and all $A' \in
\fA(O')$.
\end{itemize}
The isotony assumption implies that there is a smallest $C^*$-algebra
containing all the $\fA(O)$; this will be denoted by $\fA(\bR^4)$. It
is also assumed that there exists a unit element $\eins$ in
$\fA(\bR^4)$ which is contained in all the local algebras $\fA(O)$.
Suggested by the assumptions (I) and (II), the hermitean elements in
$\fA(O)$ should be viewed as the observables of the quantum system
which can be measured at times and locations within the spacetime
region $O$. The locality (or microcausality) assumption then says that
there are no uncertainty relations between measurements carried out at
spacetime events that are spacelike with respect to each other, or
that the corresponding observables are ``jointly measurable''. In this
way, the relativistic requirement of finite propagation speed of all
effects is built into the description of a system. (See also \cite{BuSum}
for a very recent discussion of locality aspects in quantum field theory.)

Nevertheless, there is usually in quantum field theory
an abundance of states which are ``non-local''
in the sense that there are correlations between measurements carried
out in spacelike separated regions on these states which are of
quantum nature, i.e.\ there is entanglement over spacelike separations
for such states.

Given a state $\o$ on $\fA(\bR^4)$, one can associate with it a
net of ``local von Neumann algebras'' $\{\cR_{\o}(O)\}_{O \subset \bR^4}$
in the GNS-representation by setting
$$ \cR_{\o}(O) = \p_{\o}(\fA(O))''\,,$$
where $(\pi_\omega,\cH_\omega,\Omega_\omega)$ is the GNS representation
of $\omega$ (cf.\ footnote in Sec.\ 6).
On the right hand side we read the von Neumann algebra generated
by the set of operators $\p_{\o}(\fA(O)) \subset \Bof{\cH_{\o}}$.

At this point we ought to address a point which often causes
confusion. Although in the GNS-representation the state $\o$ is
given by a vector state, it need not hold that $\o$ is a pure
state for the simple reason that $\cR_{\o}(\bR^4)$ need not
coincide with $\Bof{\cH_{\o}}$, and in that case $\o$ corresponds
to the vector state $\langle \O_{\o}|\,.\,|\O_{\o}\rangle$ {\it
restricted} to $\cR_{\o}(\bR^4)$. However, restrictions of vector
states onto proper subalgebras of $\Bof{\cH_{\o}}$ are in general
mixed states.

It is very convenient to distinguish certain states by properties of
their GNS-representations. We call a state {\it covariant} if there
exists a (strongly continuous) unitary group $\{U_{\o}(a)\}_{a \in
  \bR^4}$ which in the GNS-representation acts like the translation
group:
$$ U_{\o}(a)\cA_{\o}(O) U_{\o}(a)^{-1} = \cR_{\o}(O +a)\,, $$
for all $a \in \bR^4$ and all bounded open regions $O$. Among the
class of covariant states there are two particulary important
subclasses:
\\[6pt]
{\it Vacuum states: } $\o$ is called a vacuum state if
$U_{\o}(a)\O_{\o} = \O_{\o}$ (the state is translation-invariant) and
the joint spectrum of the selfadjoint generators $P_{\mu}$, $\mu = 0,1,2,3$, of
$U_{\o}(a) = {\rm e}^{i\sum_{\mu}a^{\m}P_{\m}}$ is
contained in the closed forward lightcone $\overline{V}_+ = \{x =
(x^{\mu}) \in \bR^4 : x^0 \ge 0, \ (x^0)^2 - (x^1)^2 - (x^2)^2 -
(x^3)^2 \ge 0\}$. In other words, the energy is positive in any
inertial Lorentz frame.
\\[6pt]
{\it Thermal equilibrium states: }
 $\o$ is called a thermal equilibrium state at inverse temperature $\b
 > 0$ (corresponding to the temperature $T = 1/k\b$ where $k$ denotes
 Boltzmann's constant) if there exists a time-like unit vector $e \in
 \bR^4$, playing the role of a distinguished time axis, so that
 $U_{\o}(t \cdot e)\O_{\o} = \O_{\o}$ and
\begin{equation}
\label{KMS}
  \langle \O_{\o}| A {\rm e}^{-\b H_{\b}}B|\O_{\o} \rangle = \langle \O_{\o}
  |BA | \O_{\o} \rangle
\end{equation}
holds for (a suitable dense subset of)
 $A,B \in \cR_{\o}(\bR^4)$, where the selfadjoint operator
 $H_{\b}$ is the generator of the
time-translations in the time-direction determined by $e$, i.e.\
$U_{\o}(t\cdot e) = {\rm e}^{itH_{\b}}$, $t \in \bR$.
\\[6pt]
We should note that \eqref{KMS} is a slightly sloppy way of expressing
the condition of thermal equilibrium at inverse temperature $\b$ which
in a mathematically more precise form would be given in terms of the
so-called ``KMS boundary condition'' that refers to analyticity
conditions of the functions $t \mapsto \langle \O_{\o}|AU_{\o}(t\cdot
e)B|\O_{\o}\rangle$ (see any of the references
\cite{BratRob2,Emch,Haag,HHW,Sew} for a precise statement of the KMS boundary
condition). That way of characterizing thermal equilibrium states
has the advantage of circumventing the difficulty that ${\rm e}^{-\b
  H_{\b}}$ will usually be unbounded since the ``thermal Hamiltonian''
$H_{\b}$ in the
GNS-representation of a thermal state has a symmetric spectrum (much
in contrast to the Hamiltonians in a vacuum-state representation). We
will not enter into further details here and refer the reader to
\cite{BratRob2,Emch,Sew} for discussion of these matters. There is,
however, a point which is worth focussing attention on. The condition
of thermal equilibrium makes reference to a single direction of time, and it is
known that if a state is a thermal equilibrium state with respect to a
certain time axis $e$, then in general it won't be a thermal
equilibrium state (at any inverse temperature) with respect to another
time-direction $e'$ \cite{Narnh,Ojima}. Nevertheless, it has been
shown by J.\ Bros and D.\ Buchholz that in a relativistic
quantum field theory, the correlation functions
of a thermal equilibrium state $\o$ (with
respect to an arbitrarily given time-direction) possess, under very
general conditions, a certain
analyticity property which is Lorentz-covariant, and stronger than the thermal
equilibrium condition with respect to the given time-direction itself
\cite{BrBu}. This analyticity condition is called ``relativistic
KMS-condition''. Let us state the relativistic spectrum condition of \cite{BrBu}
in precise terms (mainly for the sake of completeness; we won't make use of
it in the following): \\[4pt]
A state $\o$ on $\fA(\bR^4)$ is said to fulfill the {\it relativistic KMS condition}
at inverse temperature $\b >0$ if $\o$ is covariant and if there exists a timelike
vector $e$ in $V_+$ (the open interior of $\overline{V}_+$) having unit
Minkowskian length, so that for each pair of operators $A,B \in \pi_{\o}(\fA(\bR^4))$
there is a function $F = F_{AB}$ which is analytic in the domain $\cT_{\b e} =
\{ z \in \bC^4 : {\rm Im}\,z \in V_+ \cap (\b e - V_+) \}$, and continuous at the
boundary sets determined by ${\rm Im}\, z =0$, ${\rm Im}\,z = \b e$ with the boundary values
$F(x) = \langle \O_\o | A U_\o(x) B | \O_\o\rangle$, $F(x + i\b e) = \langle \O_\o | B U_\o(-x)
A|\O_\o\rangle$ for $x \in \bR^4$.
\\[4pt]
We will give an indication of the nature
of those general conditions leading to the relativistic KMS-condition
since that gives us opportunity of also
introducing the lacking bits of terminology for eventually formulating
our result.

Let us start with a vacuum state $\o = \o_{\rm vac}$, and denote the
corresponding GNS-representation by
$(\pi_{\rm vac},\cH_{\rm vac},\O_{\rm vac})$ and the local von Neumann algebras
in the vacuum representation by $\cR_{\rm vac}(O)$. When one deals
with quantum fields $\boldsymbol{\phi}$ of the Wightman type, then
$\cR_{\rm vac}(O)$ is generated by quantum field operators
$\boldsymbol{\phi}(f)$ smeared with test-functions $f$ having support in
$O$. More precisely, $\cR_{\rm vac}(O) = \{{\rm
  e}^{i\boldsymbol{\phi}(f)}\,, \ {\rm supp}\,f \subset O\}''$. This is
the typical way how local algebras of observables arise in quantum
field theory. We note that in this case, the net $\{\cR_{\rm
  vac}(O)\}_{O \subset \bR^4}$ of von Neumann algebras fulfills the
condition of {\it additivity} which requires that $\cR_{\rm vac}(O)$ is
contained in $\{\cR_{\rm vac}(O_n)\,, n \in \bN\}''$ whenever the
sequence of regions $\{O_n\}_{n \in \bN}$ covers $O$, i.e.\ $O \subset
 \bigcup_{n}O_n$. The additivity requirement can therefore
be taken for granted in quantum field theory.

Now it is clear that the vacuum state $\o_{\rm vac}$ (like any state)
determines a further class of states $\o'$ on $\fA(\bR^4)$, namely
those states which arise via density matrices in its
GNS-representation:
$$ \o'(A) = {\rm Tr}(\r'\p_{\rm vac}(A))\ \ \forall \ A \in \fA(\bR^4)
$$
for some density matrix $\r'$ on $\cH_{\rm vac}$. These states are
called {\it normal} states (in the vacuum representation, in this case), and
they correspond in an obvious manner to normal states on $\cR_{\rm
  vac}(\bR^4)$. Such normal states in the vacuum representation may be
regarded as states with a finite number of particles.

For quantum systems with a finite number of degrees of freedom one
would write a thermal equilibrium state $\o_{\b}$ as a Gibbs state
$$ \o_{\b}(A) = {\rm Tr}({\rm e}^{-\b H_{\rm vac}}\p_{\rm vac}(A))\,,$$
but for a system situated in the unboundedly extended Minkowski
spacetime, ${\rm e}^{-\b H_{\rm vac}}$ won't be a density matrix since
the spectrum of the vacuum Hamiltonian $H_{\rm vac}$ will usually be
continuous. So a thermal equilibrium state is not a normal state in
the vacuum representation. What one can however do is to approximate
$\o_{\b}$ by a sequence of ``local Gibbs states''
$$ \o_{\b}^{(N)}(A) = {\rm Tr}({\rm e}^{-\b H_{\rm vac}^{(N)}}\p_{\rm
vac}(A))\,, \quad A \in \fA(O_N)\,,$$
which are restricted to bounded spacetime regions $O_N$ with suitable
local Hamiltonians $H_{\rm vac}^{(N)}$. Now one lets $O_N
\nearrow \bR^4$ as $N \nearrow \infty$, and under fairly
general assumptions on the behaviour of the theory in the vacuum
representation that are expected to hold for all physically
relevant quantum fields, it can be shown that in the limit one gets a
thermal equilibrium state $\o_{\b}$ (this is a long known result due
to Haag, Hugenholtz and Winnink \cite{HHW}) and that, moreover,  remnants of
the spectrum condition in the vacuum representation survive the limit
to the effect that the limiting state $\o_{\b}$ satisfies the
relativistic KMS-condition \cite{BrBu}.

The relativistic KMS condition has proved useful in establishing
the Reeh-Schlieder theorem for thermal equilibrium states. We shall,
for the sake of completeness, quote the relevant results in the form of a
theorem.
\begin{Theorem} \label{resultsflat}{\rm \cite{ReSchl,Jae1,DixM}}
{\it Let $\o$ be either a vacuum state on $\fA(\bR^4)$, or a thermal
equilibrium state on $\fA(\bR^4)$ satisfying the relativistic
KMS-condition. Assume also that the net $\{\cR_{\o}(O)\}_{O \subset
  \bR^4}$ fulfills additivity and that $\cH_\o$ is separable. Then it holds that:
\begin{itemize}
 \item[(a)] The set $\cR_{\o}(O)\O_{\o}$ is dense in $\cH_{\o}$, i.e.\
   the Reeh-Schlieder property holds for  $\o =
   \langle \O_{\o}|\,.\,|\O_{\o}\rangle$ with respect to $\cR_\o(O)$, whenever $O$ is an open
   region\footnote{{\rm Here and in the following, we always assume that the open set $O$ is non-void.}}.
\item[(b)] Moreover, there is a dense  set of vectors $\chi \in
  \cH_{\o}$ so that, for each such $\chi$, $\cR_{\o}(O)\chi$
   is dense in $\cH_{\o}$ for all
  open regions $O$.
\end{itemize}}
\end{Theorem}
The proof of {\it (a)}  in the vacuum case has been
given in \cite{ReSchl}. For the case of
thermal equilibrium states, a proof of this property was only
recently established by C.D. J\"akel in \cite{Jae1}. Statement {\it (b)} is
implied by {\it (a)}, as has been shown in \cite{DixM}.
We should also like to point out that the Schlieder property mentioned in the
the proof of Thm.\ \ref{RS-S-D} holds for the state $\o$, cf.\ \cite{Schl,Jae2}.

 These quoted
results in combination with Thm.\ \ref{RS-S-D} now yield:
\begin{Theorem} \label{QFTdistill}
{\it Let $\cA = \cR_{\o}(O_{\rm A})$ and $\cB = \cR_{\o}(O_{\rm B})$ be a pair
  of local von Neumann algebras of a quantum field theory\footnote{The quantum field theory
is supposed to be non-trivial in the sense that  its local observable algebras are non-abelian, and
this is also to hold for the local von Neumann algebras in the representations considered. This
is the generic case in quantum field theory and holds for all investigated quatum field
models.} in the
  representation of a state $\o$ which is either a vacuum state, or a
  thermal equilibrium state satisfying the relativistic
  KMS-condition (with $\cH_\o$ separable).

 If the open regions $O_{\rm A}$ and $O_{\rm B}$ are spacelike separated by a
 non-zero spacelike distance, then the state $\o = \langle
 \O_{\o}|\,.\,|\O_{\o}\rangle$ is 1-distillable on the bipartite
system $(\cA,\cB)$. Moreover, there is a dense set $\cX \subset \cH_{\o}$
 so that the vector states $\langle \chi|\,.\,|\chi\rangle$ are
 1-distillable on $(\cA,\cB)$ for all $\chi \in \cX$, $||\chi|| =1$. Also, $\cX$ may be
 chosen independently of $O_{\rm A}$ and $O_{\rm B}$. Consequently, the set of
 vector  states on $\cR = (\cA \cup \cB)''$ which are 1-distillable
 on $(\cA,\cB)$ is strongly dense
 in the set of all vector states.}
\end{Theorem}
\noindent {\bf Remarks}. (i) Actually, the statement of Thm.\
\ref{QFTdistill} shows distillability not only for a dense set of
vector states on $\cR$ but even for a dense set of normal states
(i.e., density matrix states) on $\cR$. To see this note that,
owing to the assumption that the spacetime regions $O_{\rm A}$ and
$O_{\rm B}$ are spacelike separated by a finite distance, there is
for $\cR$ a separating vector in $\cH_{\o}$,
 since $\O_\o$ has
just this property: There is an open region $O$ lying spacelike to
$O_{\rm A}$ and $O_{\rm B}$. By the Reeh-Schlieder property,
$\cR_{\o}(O)\O_\o$ is dense in $\cH_\o$, and hence,  $\O_\o$ is a separating
vector for $\cR \subset \cR_\o(O)'$. This implies by
Thm.\ 7.3.8 of \cite{KadRin} that, whenever $\tilde{\o}$ is a
density matrix state on $\cR$, there is a unit vector $\chi \in
\cH_{\o}$ so that $\tilde{\o} = \o_\chi | \cR$. In other words,
under the given assumptions every normal state on $\cR$ coincides
with the restriction of a suitable vector state.
\\[4pt]
(ii) It should also be noted that, under very general conditions,
vacuum representations and also thermal equilibrium
representations of quantum field theories fulfill the so-called
``split property'' (an independence property, cf.\ \cite{Haag,SJS-RMP2,Wern}), which implies
(under the conditions of Thm.\ \ref{QFTdistill}) that there exists
an abundance of normal states which are separable and even ppt on $(\cA,\cB)$
for bounded, spacelike separated regions $O_{\rm A}$ and $O_{\rm B}$.\\[4pt]
(iii) The second part of the statement, asserting that in the GNS representation
of $\o$ there is a dense set of normal states which are distillable over
causally separated regions, is closely related to a result by Clifton and Halvorson \cite{HalCl}
who show (for a vacuum state $\o$; see \cite{Jae3} for
a generalization of the argument to states satisfying the relativistic KMS condition)
 that there is a dense set of normal states in the
GNS representation of $\o$ which are Bell-correlated over spacelike separated regions.
However, they cannot deduce that Bell-correlations over spacelike separated regions
are present for the state $\o$ itself (or for a specific class of states, like those
having the Reeh-Schlieder property, which can often be constructed out of other states).
It is here where our result provides some additional information.
\\[4pt]
(iv)
In an interesting recent paper, Reznik, Retzker and Silman \cite{RRS} propose a different
method towards qualifying the degree of entanglement of a (free) quantum field vacuum
state over spacelike separated regions. Their idea is to couple each local algebra $\cA = \cR(O_{\rm A})$
and $\cB = \cR(O_{\rm B})$ to an ``external'' algebra $B(\bC^2)$. They introduce a time-dependent
coupling between the quantum field degrees of freedom in $O_{\rm A}$ and $O_{\rm B}$ and the
corresponding ``external'' algebras, which are hence supposed to represent detection
devices for quantum field excitations. It is then shown in \cite{RRS} that this dynamical
coupling, turned on for a finite amount of time during which the quantum field degrees of
freedom remain causally separated, yields an entangled partial state for the pair of detector
systems from an initially uncorrelated state coupled to the quantum field vacuum. Further
local filtering operations are then used to distill that partial detector state to an
approximate singlet state. It should, however, be remarked that the authors of \cite{RRS}
do not demonstrate the existence of Bell-correlations in the vacuum state over arbitrarily
spacelike separated and arbitrarily small spacetime regions in the sense of  \cite{Lan1,Lan2,SumWer,HalCl}, i.e.\
in the sense of proving a violation of the CHSH inequalities by the quantum field observables
themselves. Nevertheless, the approach of \cite{RRS}, while apparently less general
than the one presented here, has some interesting aspects since potentially it may allow a more
quantitative description of distillability in quantum field systems.

\section{Distillability Beyond Spacetime Horizons}
\setcounter{equation}{0}
It is worth pointing out that in Thm.\ \ref{QFTdistill} the spacelike separated regions $O_{\rm A}$ and $O_{\rm B}$
are the localization regions of the operations that Alice and Bob can apply to a given, shared state.
The spacetime pattern of any form of classical communication between Alice and Bob that might
be necessary to ``post-select'' a sub-ensemble of higher entanglement (i.e.\ to normalize the
state $\o^{[T]}$) from a given shared ensemble
(on which local operations have been applied) is not represented in the criterion of distillability.
Put differently, the distillability criterion merely tests if there are sufficiently ``non-classical''
long-range correlations in the shared state $\o$ which can be provoked by local operations. It does not
require that the post-selection is actually carried out via classical communication realizable between
Alice and Bob in spacetime.
Such a stronger demand would have to make reference to the causal structure of the spacetime into which
Alice and Bob are placed.

We will illustrate this in the present section, and we begin by noting that Thm.\ \ref{QFTdistill} can actually
be generalized to curved spacetime. Thus, we assume that $M$ is a four-dimensional
smooth spacetime manifold, endowed with a Lorentzian metric $g$. To avoid any causal pathologies,
we will henceforth assume that $(M,g)$ is globally hyperbolic (cf.\ \cite{WaldGR}). In this case, it is possible
to construct nets of local observable ($C^*$-) algebras $\{\fA(O)\}_{O\subset M}$ for
quantized free fields,
like the scalar Klein-Gordon, Dirac and free electromagnetic fields \cite{Dimock,Waldqf}. Let us focus, for
simplicity, on the free quantized Klein-Gordon field on $(M,g)$, and denote by $\{\fA(O)\}_{O\subset M}$
 the corresponding net of local observable algebras, fulfilling the conditions
of isotony and locality, which can be naturally formulated also in
curved spacetimes.

Let us briefly indicate how the local $C^*$-algebras $\fA(O)$ are constructed in the case of the
free scalar Klein-Gordon field; for full details, see \cite{Dimock,KayWald,Waldqf}. The Klein-Gordon
operator on $(M,g)$ is $(\nabla^\mu\nabla_\mu + m^2)$ where $\nabla$ denotes the covariant derivative
of the spacetime metric $g$ and $m \ge 0$ is some constant. Owing to global hyperbolicity of
the underlying spacetime $(M,g)$, the Klein-Gordon operator possesses uniquely determined
advanced and retarded fundamental solutions (Green's functions), $G_+$ and $G_-$, which can be
viewed as distributions on $C_0^\infty(M \times M,\bR)$. Their difference $G = G_+ - G_-$ is
called the causal propagator. One can construct a $*$-algebra $\fA(M)$ generated by symbols
$W(f)$, $f \in C_0^\infty(M,\bR)$, fulfilling the relations $W(f_1)W(f_2) = {\rm e}^{-iG(f_1,f_2)/2}
W(f_1 + f_2)$, $W(f)^* = W(-f)$ and $W(f + (\nabla^\mu\nabla_\mu + m^2)h) = W(f)$. 
This algebra possesses a unit element and admits a unique
$C^*$-norm. We identify $\fA(M)$ with the $C^*$-algebra generated by all the $W(f)$. Then
$\fA(O)$ is defined as the $C^*$-subalgebra generated by all $W(f)$ where $f \in C_0^\infty(O,\bR)$.

Now, unless $(M,g)$ possesses time-symmetries, there are no obvious criteria to single out
vacuum states or thermal equilibrium states on $\fA(M)$. Nevertheless, there is a class of
preferred states on $\fA(M)$ which serve, for most purposes, as replacements for vacua
or thermal equilibrium states. The states in this class are called quasifree Hadamard states.
Given such a state, $\o$, one has $\pi_\o(W(f)) = {\rm e}^{i\Phi_\o(f)}$ in the GNS representation
of $\o$ with selfadjoint quantum field operators $\Phi_\o(f)$ in $\cH_\o$ depending linearly on
$f$ and fulfilling $\Phi_\o((\nabla^\mu\nabla_\mu + m^2)f) = 0$ and the canonical commutation
relations in the form $[\Phi_\o(f_1),\Phi_\o(f_2)] = iG(f_1,f_2)\eins$. The Hadamard condition is
a condition on the two-point distribution $\langle \O_\o|\Phi_\o(x)\Phi_\o(y)|\O_\o\rangle$
of $\o$ (symbolically written as integral kernel with $x,y \in M$) and demands, essentially,
that this has a leading singularity of the type ``1/(squared geodesic distance between $x$ and $y$)''.

Quasifree Hadamard states are a very well investigated class of the free scalar field in curved
spacetime. The reasons why they are considered as replacements for vacuum states or thermal
equilibrium states are discussed, e.g., in the refs.\ \cite{Fulling,KayWald,Waldqf,FewVer}.

The Hadamard condition on the two-point distribution of a (quasifree) state $\o$ can equivalently
be expressed by requiring that the $C^\infty$-wavefront set of the Hilbert-space valued
distribution $C_0^\infty(M) \owns f \mapsto \Phi_\o(f)\O_\o$ is confined to the set of
future-pointing causal covectors on $M$ (cf.\ \cite{SVW} and also refs.\ cited there).
If $\o$ satisfies this latter condition, one says that it fulfills the 
{\it microlocal spectrum condition} ($\mu SC$).
If the latter condition holds even with the analytic wavefront set in place of the $C^\infty$-wavefront
set, then one says that $\o$ fulfills the {\it analytic microlocal spectrum condition} ($a \mu SC$)
\cite{SVW}. (For $a \mu SC$, it is also required that the spacetime $(M,g)$ be real analytic.)
While the definitions of $C^\infty$-wavefront set and analytic wavefront set are a bit involved
so that we do not present them here and refer to \cite{SVW} and refs.\ given there for full details,
we put on record that for any quasifree state $\o$ on the obervable algebra $\fA(M)$ of the
scalar Klein-Gordon field one has
$$ \o\ \ {\rm fulfills}\ \ a \mu SC \ \ \Rightarrow\ \ \o\ \ {\rm fulfills}\ \ \mu SC \ \ \Leftrightarrow\ \
 \o \ \ {\rm Hadamard}\,.$$
Moreover, on a stationary, real analytic, globally hyperbolic spacetime $(M,g)$,
the quasifree ground states or quasifree thermal equilibrium states on $\fA(M)$, which are
known to exist under a wide range of conditions, fulfill $a \mu SC$ \cite{SVW}.
It is also known that there exist very many quasifree Hadamard states on $\fA(M)$ for
any globally hyperbolic spacetime $(M,g)$.

Several properties of the local von Neumann algebras $\cR_\o(O)$ are known for quasifree Hadamard states
$\o$, and we collect those of interest for the present discussion in the following
Proposition.
\begin{Proposition} \label{ResultsCST}
Let $(M,g)$ be a globally hyperbolic spacetime, and let $\o$ be a quasifree Hadamard state on
$\fA(M)$, the algebra of observables of the Klein-Gordon field on $(M,g)$. Write $\cR_\o(O) =
\pi_\o(\fA(O))''$, $O \subset M$, for the local von Neumann algebras in the GNS representation
of $\o$. Then the following statements hold.
\begin{itemize}
\item[(a)] $\cR_\o(O)$ is non-abelian whenever $O$ is open.
\item[(b)] There is a dense set of vectors $\chi \in \cH_\o$ so that, for each such $\chi$,
$\cR_\o(O)\chi$ is dense in $\cH_\o$, for all open $O \subset M$.
\item[(c)] If $(M,g)$ is real analytic and if $\o$ satisfies the $a \mu SC$, then the Reeh-Schlieder
property holds for $\o = \langle \O_\o|\,.\,|\O_\o\rangle$ with respect to $\cR_\o(O)$, whenever
$O \subset M$ is open.
\end{itemize}
\end{Proposition}

{\it Proof.} Statement $(a)$ is clear from the fact that the canonical commutation relations
hold for the field operators $\Phi_\o(f)$. Statement $(c)$ is a direct consequence of Thm.\ 5.4 in
\cite{SVW}. For statement $(b)$, one can argue as follows. For a globally hyperbolic
$(M,g)$, there is a countable neighbourhood base $\{O_n\}_{n\in\bN}$ for the topology of $M$
where each $O_n$ has a special shape (called ``regular diamond'' in \cite{VerSC}; we assume here also
that each $O_n$ has a non-void causal complement), which allows the conclusion that
each $\cR_\o(O_n)$ is a type ${\rm III}_1$ factor (cf.\ Thm.\ 3.6 in \cite{VerSC}).
Since $\cH_\o$ is separable (cf.\ again Thm.\ 3.6 in \cite{VerSC}), one can make use of Cor.\ 2 and Prop.\ 3
of \cite{DixM} which leads to the conclusion that there is a dense set $\cX \subset \cH_\o$
so that each $\chi \in \cX$ is cyclic for all $\cR_\o(O_n)$, $n \in \bN$. Since $\{O_n\}_{n\in\bN}$
is a neighbourhood base for the topology of $M$, each open set $O \subset M$ has $O_n \subset O$
for some $n$, and hence each $\chi \in \cX$ is cyclic for $\cR_\o(O)$ whenever $O$ is an open subset of
$M$. \qed

As in the previous section, we can conclude distillability from the just asserted Reeh-Schlieder
properties.
\begin{Theorem} \label{CSTdistill}
Let $(M,g)$ be globally hyperbolic spacetime, and
let $\o$ be a quasifree state on the observable algebra $\fA(M)$ of the
quantized scalar Klein-Gordon field on $(M,g)$.
Let $O_{\rm A}$ and
$O_{\rm B}$ be two open subsets of $M$ whose closures are causally
separated (i.e., they cannot be connected by any causal curve),
 and let $\cA = \cR_\o(O_{\rm A})$, $\cB = \cR_\o(O_{\rm
B})$.
The following statements hold:
\begin{itemize}
\item[(a)] If $(M,g)$ is real analytic and $\o$ satisfies the $a \mu SC$, then
the state $\o = \langle \O_\o |\,.\,|\O_\o\rangle$ is
1-distillable on $(\cA,\cB)$.
\item[(b)] There is a dense set $\cX \subset \cH_{\o}$
 so that the vector states $\langle \chi|\,.\,|\chi\rangle$ are
 1-distillable on $(\cA,\cB)$ for all $\chi \in \cX$, $||\chi|| =1$. Also, $\cX$ may be
 chosen independently of $O_{\rm A}$ and $O_{\rm B}$. Consequently, the set of
 normal  states on $\cR = (\cA \cup \cB)''$ which are 1-distillable
 on $(\cA,\cB)$ is strongly dense in the set of normal states on $\cR$.
\end{itemize}
\end{Theorem}
The proof of this Theorem is a straightforward combination of the statements
of Prop.\ \ref{ResultsCST} with Thm.\ \ref{RS-S-D}. For part $(b)$, we have
already made use of the observation of Remark (i) following Thm.\ \ref{QFTdistill}.

Again, as noted in Remark (iii) following Thm.\ \ref{QFTdistill}, part $(b)$ of the
last Theorem is related to a similar statement by Clifton and Halvorson \cite{HalCl}
which refers to the existence of a dense set of normal states which are Bell-correlated
over causally separated spacetime regions. Also here, our comments of Remark (iii) apply.

In Thm.\ \ref{CSTdistill} the localization regions $O_{\rm A}$ and
$O_{\rm B}$ of the system parts controlled by Alice and Bob,
respectively, could also be separated by spacetime horizons. Let
us give a concrete example and take $(M,g)$ to be
Schwarzschild-Kruskal spacetime, i.e.\ the maximal analytic
extension of Schwarzschild spacetime. This is a globally
hyperbolic spacetime which is real analytic, and it has two
subregions, denoted by {\bf I} and {\bf II}, that model the
interior and exterior spacetime parts of an eternal black hole,
respectively (see Sec.~6.4 in \cite{WaldGR}). These two regions
are separated by the black hole horizon, so that no classical
signal can be sent from the interior region {\bf I} to an observer
situated in the exterior region {\bf II}. The situation is
depicted in Figure~2 below.

For the quantized scalar Klein-Gordon field on the
Schwarzschild-Kruskal spacetime, there is a preferred quasifree
state, the so-called Hartle-Hawking state, which is in a sense the
best candidate for the physical ``vacuum'' state on this spacetime
(cf.\ \cite{KayDW,Waldqf}). It is generally believed that this
state fulfills the $a \mu SC$ on all of $M$. (The arguments of
\cite{SVW} can be used to show that $a \mu SC$ is fulfilled in
region {\bf II} and its ``opposite'' region, which makes it
plausible that this holds actually on all of $M$, although there
is as yet no complete proof.) Anticipating that this is the case,
we can choose the localization region $O_{\rm A}$ inside the
interior region {\bf I} and $O_{\rm B}$ in the exterior region
{\bf II} (cf.\ Fig.\ 2). Then, by our last theorem, we find that the
Hartle-Hawking state $\o$ of the quantized Klein-Gordon field is
distillable on the bipartite system $(\cA,\cB)$ with
$\cA = \cR_\o(O_{\rm A})$ and $\cB = \cR_\o(O_{\rm B})$.
Furthermore, there is a dense set of normal states in the
GNS-representation of the Hartle-Hawking state with respect to
which this distillability holds. (At any rate, since the existence
of quasifree Hadamard states for the Klein-Gordon field on
the Schwarzschild-Kruskal spacetime is guaranteed, part (b) of Thm.\
\ref{CSTdistill} always ensures the existence of an abundance of states
which are distillable on $(\cA,\cB)$).

A similar example for regions $O_{\rm A}$ and $O_{\rm B}$
separated by a spacetime horizon (an event horizon) can be given
for de Sitter spacetime; the de Sitter ``vacuum state'' for the
quantized Klein-Gordon field actually has all the required
properties for the distillablity statement of Thm.\
\ref{CSTdistill}, cf.\ \cite{BrosMosch}.

This shows that distillability of quantum field states beyond
spacetime horizons (event horizons) can be expected quite
generally.

A similar situation occurs also in the standard Friedmann-Robertson-Walker
cosmological models with an initial spacetime singularity. In this
scenario, spacetime regions sufficiently far apart from each other
are causally separated  for a finite amount of time by their
cosmological horizons \cite{WaldGR}. However, also in this situation,
a quantum field state fulfilling the $a \mu SC$ on any Friedmann-Robertson-Walker
spacetime would be distillable on a bipartite system $(\cA,\cB)$ of the form
$\cA = \cR_\o(O_{\rm A})$ and $\cB = \cR_\o(O_{\rm B})$ for spacetime regions
$O_A$ and $O_B$ separated by a cosmological horizon. Again, there is at any
rate a large class of states where such a distillability is found.
In passing we should like to note that quantum field correlations, whose appearance is
precisely expressed by the Reeh-Schlieder property, have already been considered
in connection with the question if (potentially, very strong) quantum field fluctuations in the early universe
could account for the structure of its later development \cite{WaldCBH}.
\\[6pt]
\begin{center}
\epsfig{file=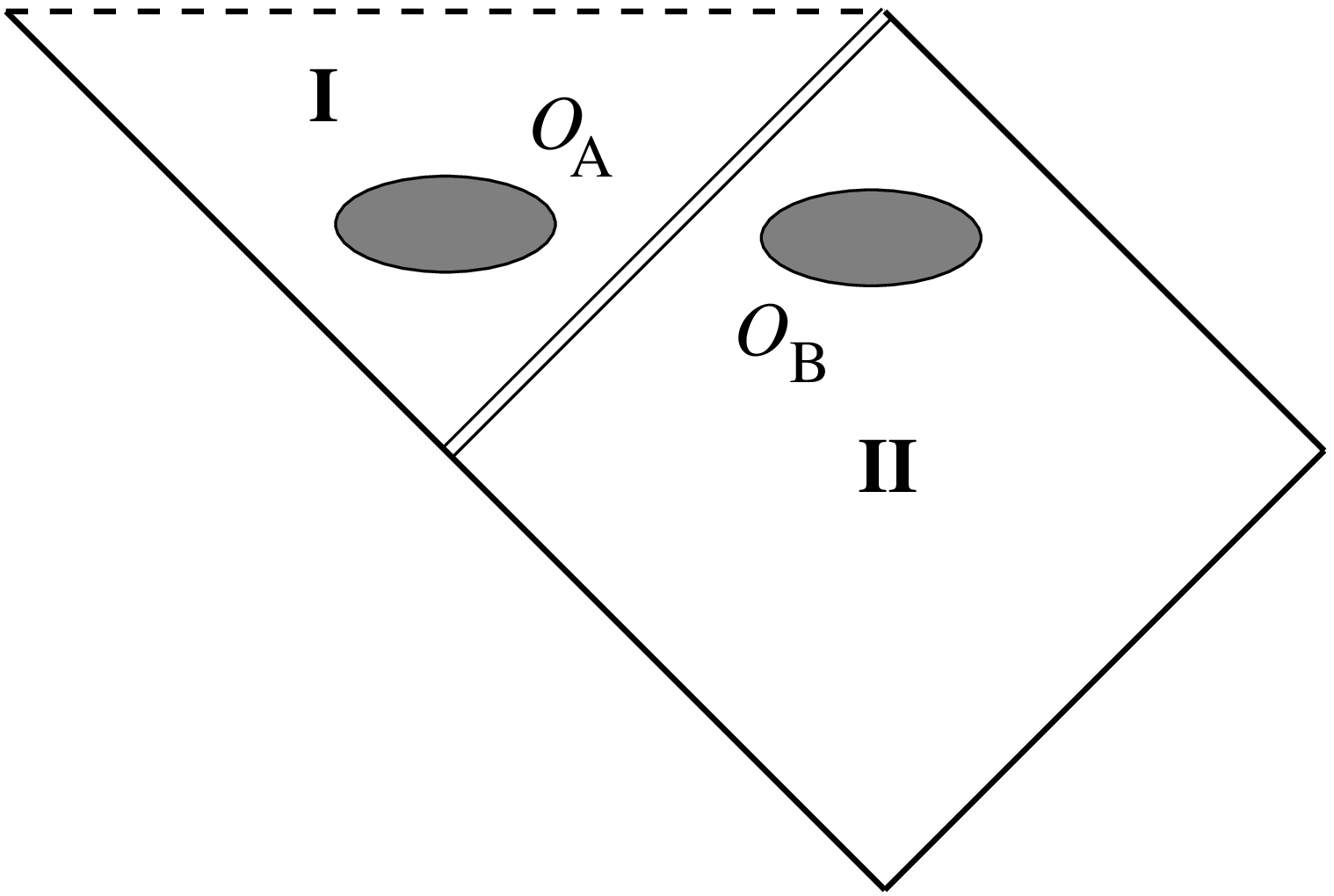,width=8.5cm}
\end{center}
{\small {\bf Figure 2}. This figure shows the interior region {\bf
I} and exterior region {\bf II} of the conformal diagram of
 Schwarzschild-Kruskal spacetime, which is a
model of a static black hole spacetime (at large times after collapse of
a star to a black hole). The event horizon, represented by the double lines,
separates region {\bf I} from region {\bf II} such that no signal can be sent
from {\bf I} to {\bf II} across the horizon.
 A quantum field state which satisfies the Reeh-Schlieder property (as e.g.\
implied by the analytic microlocal spectrum condition) is distillable over
the shaded spacetime regions $O_{\rm A}$ (wherein `Alice' conducts her
experiments on the state) and $O_{\rm B}$ (wherein `Bob' conducts his experiments
on the state). The dashed line represents the black hole singularity.}

\section{Discussion:\\ Classical Communication in Spacetime?}

Distillation was introduced as the process of taking imperfectly
entangled systems, and turning them into a useful entanglement
resource. Any such process requires classical communication, even
though for realizing 1-distillability only a single step of
post-selection is required. It is suggestive to describe the
classical communication steps also as causal communication
processes in spacetime.

This immediately raises a problem: if the laboratories of Alice
and Bob are separated by an event horizon, they will {\it never}
be able to exchange the required signals, so in this case the
above results of the previous section might appear to be totally
useless. Several comments to this idea are in order.

\noindent 1. Event horizons are {\it global} features of a
spacetime. Hence if we are interested in what can be gained from
the local state between Alice and Bob, the future development of
the universe remains yet unknown. Since the gravitational
background is taken as ``external'' at this level of the theory,
the adopted framework, using only spacetime structure up until the
time the quantum laboratories close, never allows a decision on
whether or not postselection will be causally possible.

\noindent 2. The attempt to include the distillation process in
the spacetime description meets the following characteristic
difficulty: It becomes very hard to distinguish between classical
and quantum communication. Obviously, a quantum operation disturbs
the quantum field in its future light cone, but it is very hard to
assert that this disturbance leaves alone the spacetime region
where the negotiations for postselection take place. In other
words: we cannot distinguish LOCC operations from exchanging
quantum particles, and this would completely trivialize the
distinction between distillable and separable states.

\noindent 3. This difficulty is akin to the problem of realizing
statistical experiments in spacetime. On the one hand, the
statistical interpretation of quantum mechanics (and hence of
quantum field theory) is based on independent repetitions of ``the
same'' experiment. But in a dynamic space time it is clear that
strictly speaking no repetition is possible, and the above
disturbance argument casts additional doubt on the possibility of
{\it independent} repetitions. Carrying this argument still
further, into the domain of quantum cosmology, it has been debated
\cite{Fink} whether quantum theory may ever apply to the universe
as a whole. Whether this can be resolved by showing that for
typical (small) experimental setups statistical behavior can be
shown to hold with probability 1 in any ensemble of universes
admitted by the theory is a question far beyond the present
paper.

 To summarize: we have adopted here the most ``local''
approach to distillability, where it is strictly taken as a
property of a state $\o$ of a general bipartite quantum system
$(\cA,\cB) \subset \cR$, independent of the ``surroundings'' of that quantum system
and the global structure of the spacetime into which it is placed.
Still, it would be quite interesting to see if distillability
criteria taking into account the realizability of distillation
protocols in spacetime can be developed in a satisfactory manner
(e.g.\ reconcilable with ideas like general covariance
\cite{BFV}, and with the difficulties related to independence of measurements
alluded to above). We should finally note that the difference between
these two points of view is insignificant for present day
laboratory physics where it can always be safely assumed that
spacetime is Minkowskian.

\end{document}